\documentclass[10pt,conference,letter]{./IEEEtran}

\usepackage{graphicx,graphics,subfigure,wrapfig}
\usepackage{epsfig}
\usepackage{psfrag}
\usepackage[usenames]{color}
\usepackage{multicol,amsmath,dsfont,amssymb}
\usepackage{cite,url,fancybox,balance}
\usepackage{array,multirow}
\usepackage[active]{srcltx}
\usepackage{float}
\usepackage{bm}
\usepackage{cmll}
\usepackage{enumerate, algorithmic}
\newtheorem{theorem}{Theorem}%[section]
\newtheorem{proposition}{Proposition}%[section]
\newfont{\mbb}{msbm10 scaled 1100}

\definecolor{red}{RGB}{210,20,50}

\definecolor{lblue}{RGB}{30,160,240}

\definecolor{green}{RGB}{75,180,130}

\definecolor{blue}{RGB}{0,0,255}

\definecolor{magenta}{RGB}{255,0,255}

\definecolor{orange}{RGB}{255,128,0}

\definecolor{dgreen}{RGB}{0,151,0}

\def\b0{{\bf 0}}

\newsavebox{\ieeealgbox}

\DeclareMathOperator*{\argmax}{arg\,max}
\DeclareMathOperator*{\argmin}{arg\,min}
\begin{document}

\title{Reducibility of joint relay positioning and flow optimization problem}

\author{
\authorblockN{Mohit Thakur}
\authorblockA{Institute for Communications Engineering,\\
Technische Universit\"{a}t M\"{u}nchen,\\
80290, M\"{u}nchen, Germany.\\
mohit.thakur@tum.de}
\and
\authorblockN{Nadia Fawaz}
\authorblockA{Technicolor Research Center, \\
Palo Alto, CA, USA.\\
nadia.fawaz@technicolor.com}
\and
\authorblockN{Muriel M\'{e}dard}
\authorblockA{Research Laboratory for Electronics,\\
Massachusetts Institute of Technology,\\
Cambridge, MA, USA.\\
medard@mit.edu}
}

\vspace{-6mm}
\maketitle
\vspace{-2mm}

\begin{abstract}
This paper shows how to reduce the otherwise hard joint relay
positioning and flow optimization problem into a sequence a two
simpler decoupled problems. We consider a class of wireless
multicast hypergraphs mainly characterized by their hyperarc rate
functions, that are increasing and convex in power, and decreasing
in distance between the transmit node and the farthest end node of
the hyperarc. The set-up consists of a single multicast flow session
involving a source, multiple destinations and a relay that can be
positioned freely. The first problem formulates the relay
positioning problem in a purely geometric sense, and once the
optimal relay position is obtained the second problem addresses the
flow optimization. Furthermore, we present simple and efficient
algorithms to solve these problems.
\end{abstract}

% \begin{keywords}
% Low-SNR, broadcast relay channel, geometry.
% \end{keywords}

%------------------------------------------------------------------------%
\section{INTRODUCTION} \label{sec:Introduction}
%------------------------------------------------------------------------%
We consider a version of network planning problem under a relatively
simple construct of a single session consisting of a source $s$, a
destination set $T$ and an arbitrarily positionable relay $r$, all
on a $2$-D Euclidean plane. The problem can then be stated as:
\emph{ What is optimal relay position that maximizes the multicast
flow from $s$ to $T$?} Similarly, we can also ask: \emph{What is the
optimal relay position that minimizes the cost (in terms of total
network power) for a target multicast flow $F$?}

A fairly general class of acyclic hypergraphs are considered. The
hypergraph model is characterized by the following rules of
construction of the hypergraph $\mathcal{G(N,A)}$:
\begin{enumerate}
 \item $\mathcal{G(N,A)}$ consists of finite set
of nodes $\mathcal{N}$ positioned on on a $2$-D Euclidean plane and
a finite set of hyperarcs $\mathcal{A}$.
\item Each hyperarc in $\mathcal{A}$ emanates from a
transmit node and connects a set of receivers (or end nodes) in the
system. Also, each hyperarc is associated with a rate function that
is convex and increasing in transmit node power and decreasing in
distance between the transmit node and the farthest node spanned by
the hyperarc in the system.
\item Each end node spanned by the hyperarc can decode the information
sent over the hyperarc equally reliably, i.e. all the end nodes of
an hyperarc get equal rate.
\end{enumerate}

In relation to the special case of our hypergraph model, the authors
addressed the first question (max-flow) in the context of Low-SNR
Broadcast Relay Channel in \cite{Thakur-Fawaz-Medard-arXivISIT2011}.

This paper has two major contributions. Firstly, we solve the
general joint relay positioning and max-flow optimization problem
for our hypergraph model. Secondly, we address the min-cost flow
problem and establish a relation of duality between the max-flow and
min-cost problems. An efficient algorithm that solves the joint
relay positioning and max-flow problem is presented, in addition to
an algorithm that solves an important special case of the min-cost
problem.

The relay positioning problem has been studied in various settings
\cite{Aggarwal-Bennetan-Calderbank-2009,Ergen-Varaiya-2006,Cannons-Milstein-Zeger-2009}.
In most cases, the problem is either heuristically solved due to
inherent complexity, or approximately solved using simpler methods
but compromising accuracy. We reduce the non-convex joint problem
into easily solvable sequence of two decoupled problems. The first
problem solves for optimal relay position in a purely geometric
sense with no flow optimization involved. Upon obtaining the optimal
relay position, the second problem addresses the flow optimization.
The decoupling of the joint problem comes as a consequence of the
convexity (in power) of hyperarc rate functions.

The next section develops the wireless network model.
Section~\ref{sec:Multiflow} presents the key multicast flow
concentration ideas for max-flow and min-cost flow that are central
to the reducibility of the joint problem. In
Section~\ref{sec:Algorithms}, we present the algorithms and
Section~\ref{sec:Example} contains an example where the results of
this paper are applied. Finally, we conclude in
Section~\ref{sec:Conclusion}.

%------------------------------------------------------------------------%
\section{PRELIMINARIES AND MODEL} \label{sec:Prelim}
%------------------------------------------------------------------------%

Consider a wireless network hypergraph $\mathcal{G(N,A)}$ consisting
of $|\mathcal{N}|=n+2$ nodes placed on a $2$-D Euclidean plane with
$|\mathcal{A}|$ number of hyperarcs and the only arbitrarily
positionable node as the relay $r$. The node set
$\mathcal{N}=\{s,r,t_1,..,t_n\}$ consists of a source node $s$, a
relay $r$ and an ordered destination set $T=\{t_1,..,t_n\}$ (in
increasing distance from $s$). Their positions on the $2$-D
Euclidean plane are denoted by the set of two-tuple vector
$\mathcal{Z}=\{z_i=(x_j,y_j)|\forall j \in \mathcal{N}\}$.

\begin{figure}[tp]
\begin{center}
\psfrag{s}{$s$} \psfrag{t1}{$t_{1}$} \psfrag{t2}{$t_{2}$}
\psfrag{d}{$(d)$} \psfrag{r}{$r$} \psfrag{e}{$(e)$}
\psfrag{f}{$(f)$} \psfrag{a}{$(a)$} \psfrag{b}{$(b)$}
\psfrag{c}{$(c)$} \psfrag{t}{$t$}
\includegraphics[width=1\columnwidth]{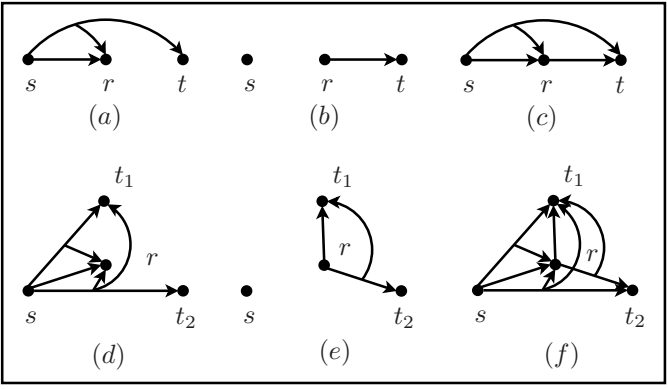}
\end{center}
\vspace{-4mm} \caption{{ Hyperarcs are constructed in increasing
order of distance from the transmitter. (a)-(c): $3$ node system.
(a): Source hyperarc set - $\{(s,r),(s,rt)\}$. (b): Relay hyperarc
set - $\{(r,t)\}$. (c): Hypergraph $\mathcal{G(N,A)}$. (d)-(f): $4$
node system with $T=\{t_1,t_2\}$ such that $D_{sr}<D_{st_1} <
D_{st_2}$ and $D_{rt_1} < D_{rt_2}$. (d) Source hyperarc set -
$\{(s,r),(s,rt_1),(s,rt_1t_2)\}$. (e) Relay hyperarc set -
$\{(r,t_1),(r,t_1t_2)\}$. (f): Hypergraph $\mathcal{G(N,A)}$.}}
\label{fig:Fig1} \vspace{-6mm}
\end{figure}

All hyperarcs in $\mathcal{A}$ are denoted by $(u,V_{k_u})$, where
$u$ is the transmit node and $V_{k_u}=\{v_1,..,v_{k_u}\}$ is the
ordered set (in increasing distance $u$) of end nodes of the
hyperarc, and $V_{k_u} \subset \mathcal{N}\backslash \{u\}$. The
hyperarcs emanating from a transmitter node are constructed in order
of increasing distances of the receivers from the transmitter (refer
Figure~\ref{fig:Fig1}). This construction rule captures the distance
based approach and is analogous to time sharing for broadcasting.
Note that, this is one technique to construct the hypergraph
$\mathcal{G(N,A)}$, our model allows arbitrary styles of hypergraph
construction that follow the above three mentioned rules. Although,
since time sharing is optimal for broadcasting we will stick to this
technique as the main example in this paper. All the nodes in the
set $V_{k_u}$ receive the information transmitted over the hyperarc
$(u,V_{k_u})$ equally reliably. Any hyperarc $(u,V_{k_u}) \in
\mathcal{A}$ is associated with a rate function
$R^u_{v_{k_u}}=f(P^u_{v_{k_u}},D_{uv_{k_u}})$, where $P^u_{k_u}$ and
$D_{uv_{k_u}}$ denotes the fraction of the total transmit node power
allocated for the hyperarc and the Euclidean distance between
transmit node $u$ and the farthest end node $v_{k_u}$, respectively.

The hyperarc rate function $R^u_{v_{k_u}}$ is increasing and convex
in power $P^u_{v_{k_u}}$ and decreasing in $D_{uv_{k_u}}$.
Furthermore, without loss of generality, we write the hyperarc rate
function into two separable functions of power and distance
\begin{equation}\label{ratefunc}
R^{u}_{v_{k_u}}=\frac{g(P^{u}_{v_{k_u}})}{h(D_{uv_{k_u}})}
\hspace{2mm} \mbox{or} \hspace{2mm}
R^{u}_{v_{k_u}}=g(P^{u}_{v_{k_u}}) - h(D_{uv_{k_u}}),
\end{equation}
where $g:\mathbf{R}^{+} \longrightarrow \mathbf{R}^{+}$ is
increasing and convex and $h:\mathbf{R}^{+} \longrightarrow
\mathbf{R}^{+}$ is increasing.
% % We assume that the functions $g$ and $h$ are commutative in the ranges, i.e. for
% % any two elements $e_1$ and $e_2$ the following holds,
% \begin{equation*}
% h(e_1)h(e_2)=h(e_2)h(e_1) \hspace{1mm} \text{and} \hspace{1mm}
% g(e_1)g(e_2)=g(e_2)g(e_1).
% \end{equation*}
Mainly, we will be concerned with the first equation in
(\ref{ratefunc}). Moreover, to comply with standard physical
wireless channel models we assume that
\begin{equation}\label{ratefuncder}
\begin{split}
 \frac{\partial{g(P^{u}_{v_{k_u}}})}{\partial{P^{u}_{v_{k_u}}}}
\leq \frac{\partial{h(D_{uv_{k_u}}})}{\partial{D_{uv_{k_u}}}},
\end{split}
\end{equation}
$\forall (P^{u}_{v_{k_u}}=D_{uv_{k_u}}) \in
\mbox{$\mathbf{dom}(P^{u}_{v_{k_u}},D_{uv_{k_u}})$}$. If the
functions $g$ and $h$ are not differentiable entirely in
$\mathbf{dom}(P^{u}_{v_{k_u}},D_{uv_{k_u}})$, then
Inequality~\ref{ratefuncder} can be rewritten with partial
sub-derivatives, implying that differentiability is not imperative.

Denote the convex hull of the nodes in $\mathcal{N}\backslash \{r\}$
by $\mathcal{C}$. For a given relay position $z_r \in \mathcal{C}$,
let $L_i=\{l^{i}_{1},..,l^{i}_{\tau_i}\}$ be the set of paths from
$s$ to a destination $t_i \in T$ and let $L=\{l_1,..,l_{\tau}\}$ be
the set of paths from $s$ that span all the destination set $T$,
therefore $L \subset \bigcup_{i \in [1,n]} L_{i}$. Moreover, any
path in the system consists of either a single hyperarc or at most
two hyperarcs as there are only two transmitters in the system. Let
$\mu$ and $\nu$ denote the total given power of source and relay,
respectively, and $\gamma=\frac{\nu}{\mu}$ denote their ratio, where
$\gamma \in (0, \infty)$. Denote with $F_{l^{i}_{j}}$ and $F_{i}$
the flow over the path $l^{i}_{j}$ (for $j \in[1,\tau_i]$) and the
total flow to the destination $t_{i} \in T$, respectively, such that
$F_{i}=\sum_{j \in[1,\tau_i]} F_{l^{i}_{j}}$.  Define $F$ to be the
the multicast flow from $s$ to the destination set $T$ as the
minimum among the total flows to each destination, then for a given
relay position $z_r \in \mathcal{C}$ the multicast max-flow problem
can be written as, \vspace{-2mm}
\begin{equation*}
\hspace{10mm} \mbox{Maximize} \hspace{10mm} \left(F =
\displaystyle\min_{i \in [1,n]} F_{i} \right) \hspace{23mm}
\mbox{(A)}
\end{equation*}\vspace{-4mm}
\begin{align}
\hspace{-4mm} \mbox{subject to:} \hspace{2mm}
&F_{i} \leq \displaystyle\sum_{j=1}^{\tau_{i}} F_{l^{i}_{j}}, \forall i \in [1,n],\\
& 0 \leq F_{l^{i}_{j}} \in \mathfrak{C}(P,D), \hspace{2mm} \forall j
\in [1,\tau_i], \forall i \in [1,n].
\end{align}
The hyperarc rate constraints and node sum-power constraints are
denoted by the set $\mathfrak{C}(P,D)$ in Program (A) for
simplicity. Program (A) in general is non-convex, as the path flow
function $F_{l^{i}_{j}}$ can be non-convex, e.g. let the path
$l^{i}_{j} \in L_i$ be $l^{i}_{j}=\{(s,V_{k_s}),(r,V_{k_r})\}$,
($l^{t_2}_{1}=\{(s,rt_1),(r,t_1t_2)\}$ in Figure~\ref{fig:Fig1}(f)),
then $F_{l^{i}_{j}}=\min(R^{s}_{v_{k_s}},R^{r}_{v_{k_r}})$.

Now we define the notion of cost for a given hyperarc rate
$R^{u}_{v_{k_u}}=\frac{g(P^{u}_{v_{k_u}})}{h(D_{uv_{k_u}})} \geq 0$.
The cost of rate $R^{u}_{v_{k_u}}$ is given by the total power
consumed by the hyperarc to achieve $R^{u}_{v_{k_u}}$
\begin{equation}\label{Sec2eq1}
P^{u}_{v_{k_u}}=g^{-1}\left( R^{u}_{v_{k_u}} h(D_{uv_{k_u}})
\right),
\end{equation}
where $g^{-1}:\mathbf{R}^{+} \longrightarrow \mathbf{R}^{+}$ is the
inverse function of $g$ that maps its range to its domain.
Therefore, the total cost of multicast flow $F$ is simply the sum of
powers of all the hypearcs in the system. Note that the function
$g^{-1}$ is increasing and concave, and if $h$ is convex then from
Inequality~(\ref{ratefuncder}), $g^{-1} \circ h$ increasing and
convex. So for a given relay position $z_r \in \mathcal{C}$, the
min-cost problem minimizing the total cost for setting up the
multicast session $(s,T)$ with a target flow $F$ can be written as,
\begin{equation*}
\hspace{10mm} \mbox{Minimize} \hspace{10mm} \left(P =
\displaystyle\sum_{(u,V_{k_u}) \in \mathcal{A}} P^{u}_{v_{k_u}}
\right) \hspace{13mm} \mbox{(B)}
\end{equation*}
\vspace{-4mm}
\begin{align}
\hspace{-4mm} \mbox{subject to:} \hspace{2mm} &F \leq F_{i}
\leq \displaystyle\sum_{j=1}^{\tau_{i}} F_{l^{i}_{j}}, \hspace{2mm} \forall i \in [1,n], \label{B1} \\
& \mathfrak{C}(P,D) \ni F_{l^{i}_{j}} \geq 0, \forall j \in
[1,\tau_i], \forall i \in [1,n]. \label{B2}
\end{align}
Constraint (\ref{B1}) makes sure that any destination $t_i \in T$
receives a minimum of flow $F$. Like in Program (A), we denote with
the set $\mathfrak{C}(P,D)$ the hyperarc rate and power constraints.

Finally, define the point $p^*$, that will be crucial in developing
algorithms in later sections, as
\begin{equation}\label{Sec2eq10}
z_{p^*}=\displaystyle\argmin_{z_p} (\max(\nu^* h(D_{z_ps}),\mu^*
\displaystyle\max_{t_i \in T} ( h(D_{z_pt_i})))),
\end{equation}
where, $\mu^*=g(\mu)$ and $\nu^*=g(\nu)$. An easy way to understand
$p^*$ is that if $\mu^*=\nu^*=1$ then $p^*$ is the circumcenter of
two or more nodes in the set $\mathcal{N} \backslash \{r\}$. Note
that the program in Equation~(\ref{Sec2eq10}) is a convex program.
Also, denote the optimal value of the objective function in
Equation~(\ref{Sec2eq10}) as $D_{p^*}$.

Hereafter, we represent with $(s,T,\mathcal{Z},\gamma)$ and
$(s,T,\mathcal{Z},\gamma,F)$ the joint relay positioning and flow
optimization problem instances that maximizes the multicast flow and
minimizes the total cost for a the target flow $F$, and with
$z^{*}_{\gamma \uparrow }$ and $z^{*}_{F \downarrow }$ denote the
optimal relay positions, respectively.

%------------------------------------------------------------------------%
\section{Multicast Flow Properties And Reduction}\label{sec:Multiflow}
%------------------------------------------------------------------------%
In this section we develop fundamental multicast flow properties
that govern the multicast flow in the wireless network hypergraphs
that we consider in this paper. First, we briefly note the main
hurdles in jointly optimizing the problem. For a given problem
instance different relay positions can result in different
hypergraphs, which makes the use of standard graph-based flow
optimization algorithms difficult. Moreover, the hyperarc rate
function can be non-convex itself.

We will show that the joint problems $(s,T,\mathcal{Z},\gamma)$ and
$(s,T,\mathcal{Z},\gamma,F)$ can be reduced to solving a sequence of
two decoupled problems. The reduced problems are decoupled in the
sense that the first problem is purely a geometric optimization
problem and involves no flow optimization and vice versa for the
second problem. At the same time, they are not entirely decoupled
because the two problems need to be solved in succession and cannot
be solved separately. Now we present a series of results that are
fundamental to the reducibility of the joint problem.

\begin{proposition}\label{P1}
 The optimal relay positions $z^{*}_{\gamma \uparrow }$
and $z^{*}_{F \downarrow }$  lie inside the convex hull
$\mathcal{C}$.
\end{proposition}

Refer Appendix~\ref{ap:Proof-P1} for the proof. Proposition~\ref{P1}
tells us that only the points inside the polygon $\mathcal{C}$ need
to be considered. This brings us to the following fundamental
theorem.
\begin{theorem}[Flow Concentration] \label{T1}
Given $z_r \in \mathcal{C}$:
\begin{enumerate}[(i)]
\item  the maximized multicast flow $F^*$
concentrates over at most two paths from $s$ to the destination set
$T$.
\item for any target flow $F \in [0,F^*]$ the min-cost multicast flow
concentrates over at most two paths from $s$ to $T$.
\end{enumerate}
\end{theorem}

The proof is detailed in Appendix~\ref{ap:Proof-T1}.
Theorem~\ref{T1} is central to the two questions we aim to answer
and reduces the complexity of joint optimization greatly by
considering only two paths instead of many. Essentially,
Theorem~\ref{T1} tells that for a given relay position $z_r \in
\mathcal{C}$, the multicast flow $F$ must go only over the paths
that span all the destination set $T$, i.e. set $L$. Furthermore,
among the paths in $L$, the maximized multicast flow $F^*$ goes over
only two paths, namely the path $\hat{l}_1=\{(s,T_1),(r,T_2)\}$ that
has the highest min-cut among all the paths through the relay $r$,
and path $\hat{l}_2=\{(s,t_1,..,t_n)=(s,T)\}$, which is the biggest
hyperarc from $s$ spanning all the destination set $T$, where $r\in
T_1$ and $T_1 \cup T_2 = T$. The same holds for the min-cost case
for a given relay position $z_r \in \mathcal{C}$. Consequently, it
is also true for the optimal relay positions $z^{*}_{\gamma \uparrow
}$ and $z^{*}_{F \downarrow }$. Hereafter, we only need to consider
the flow over paths $\hat{l}_1$ and $\hat{l}_2$ (corresponding to
the relay position in consideration).

% Similarly for the min-cost problem, the optimal min-cost
% multicast flow $F \in [0,F^*]$ flows over at most two paths, namely
% $\hat{l_2}$ and $\hat{l_3}=\{(s,T'_1),(r,T'_2)\}$, where $r \in
% T'_1$, $T'_1 \cup T'_2=T$ and the path $\hat{l_3}$ is the cheapest
% path from $s$ to $T$ through $r$ for the given relay position $z_r
% \in \mathcal{C}$.

\subsection{Max-flow Problem - $(s,T,\mathcal{Z},\gamma)$}\label{sec:Maxflow}

\begin{figure}[tp]
\begin{center}
\psfrag{s}{$s$} \psfrag{t1}{$t_{1}$} \psfrag{t2}{$t_{2}$}
\psfrag{t3}{$t_{3}$} \psfrag{r'}{$\bar{r}$} \psfrag{r}{$r$}
\psfrag{inf}{$\infty$} \psfrag{z1}{$z^{*}_{1 \uparrow }$}
\psfrag{a}{$(a)$} \psfrag{b}{$(b)$} \psfrag{0}{$0$}
\psfrag{g}{$\gamma$}
\includegraphics[width=1\columnwidth]{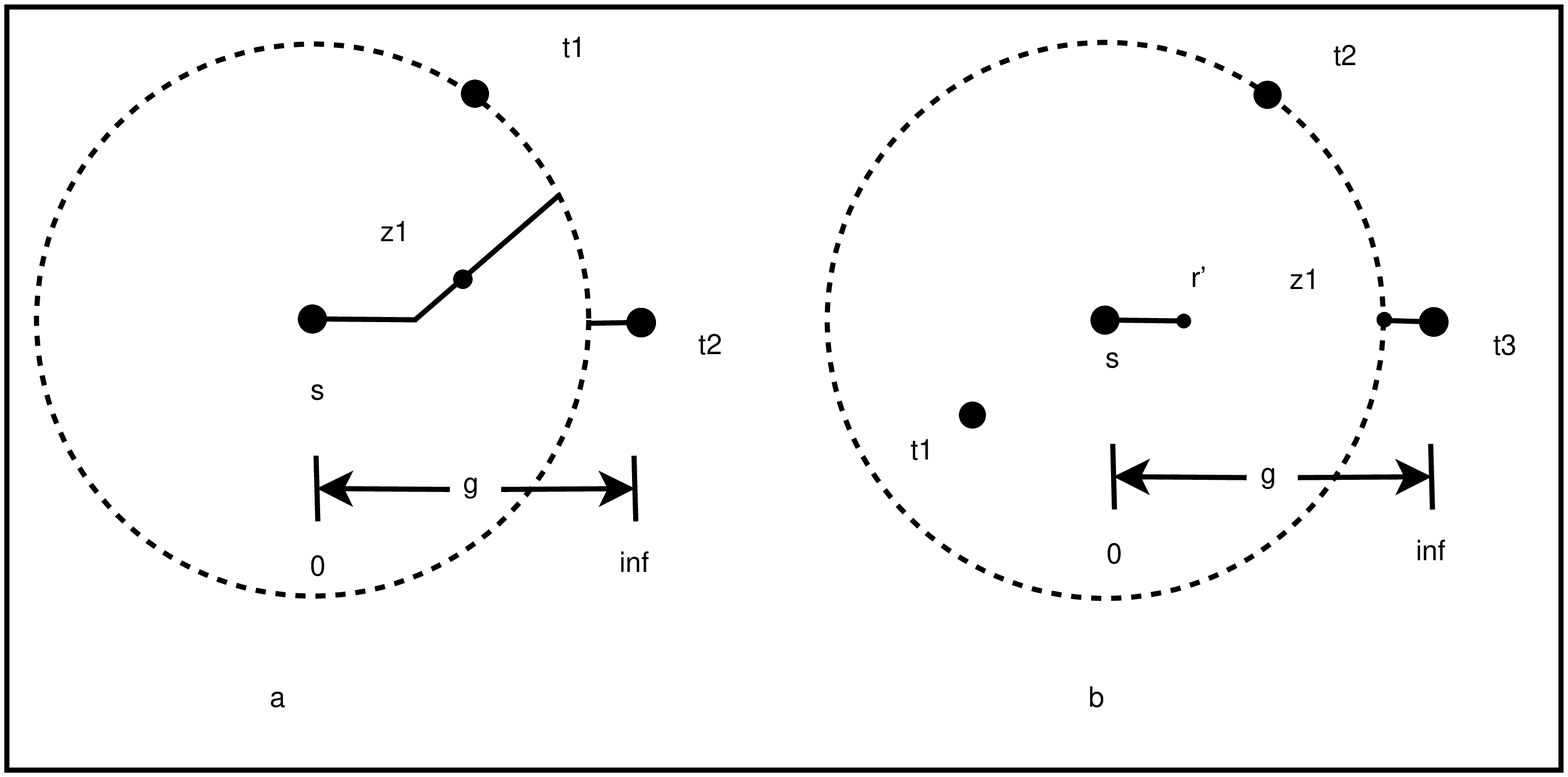}
\end{center}
\vspace{-4mm} \caption{{ The solid piecewise linear segment in
examples (a) and (b) marks the set of points $\widehat{r}$ for
different values of $\pi_s \in (0,D_{st_2})$. Each point
$\widehat{r}$ corresponds to $z^{*}_{\gamma \uparrow }$ for some
$\gamma \in (0,\infty)$. The piecewise linear segment breaks beyond
the dashed circle as $z_1 \in C^s_{T_1}$. (a): E.g. $C^{s}_r$ with
$0<\pi_s < D_{st_1}$,
$z_{\widehat{r}}=\displaystyle\argmin_{\hat{z_r} \in C^{s}_{r}}
\max(D_{\hat{z_r}t_1},D_{\hat{z_r}t_2})$. Same goes for the example
in (b).
% (b)  Here, for all $\pi_s \in
% [D_{s\bar{r}},D_{st_2})$ the point $r' =\bar{r}$, implying that
% $z_{\bar{r}}=z^{*}_{\gamma}$ solving $(s,T,\mathcal{Z},\gamma)$ for
% multiple values of $\gamma$.
}} \label{fig:Fig2} \vspace{-8mm}
\end{figure}

Assuming that the transmitted signal propagates omnidirectionally,
we can geometrically represent the hyperarcs of the path
$\hat{l}_1=\{(s,T_1),(r,T_2)\}$ by circles $C^{s}_{T_1}$ and
$C^{r}_{T_2}$ centered at $s$ and $r$ with radii $\pi_s=D_{st_k}$
and $\pi_r=D_{rt_{k'}}$ (where $D_{st_k} =\max_{t_i \in T_1}
(D_{st_i})$ and $D_{rt_{k'}} =\max_{t_j \in T_2} (D_{rt_j})$),
respectively. Similarly, the path $\hat{l_2}=\{(s,T)\}$ can be
represented by the circle $C^s_{T}$ with radius $D_{st_n}$. Also,
$C_{\cup}=C^s_{T_1} \cup C^r_{T_2}$ denotes the union region of the
two circles. Then using Theorem~\ref{T1}, Program (A) can be
re-written as,
\begin{equation*}
\hspace{4mm} \underset{\substack{P^{s}_{T_1}+P^{s}_{T} \leq \mu,
\\P^r_{T_2} \leq \nu, \pi_s, \pi_r}}{\text{Maximize}} \bigg(\min
\bigg(\frac{g({P^{s}_{T_1}})}{h(\pi_s)},\frac{g({P^{r}_{T_2}})}{h(\pi_r)}
\bigg)+\frac{g({P^{s}_{T}})}{h(D_{st_n})}  \bigg)  \hspace{5mm}
\text{(C)}
\end{equation*}
where, $P^s_{T_1}, P^r_{T_2}$ and $P^s_{T}$ are the powers for
hyperarcs of the paths $\hat{l}_1=\{C^s_{T_1}, C^r_{T_2}\}$ and
$\hat{l}_2=\{C^s_{T}\}$, respectively. The radii variables $\pi_s$
and $\pi_r$ correspond to path $\hat{l}_1$ for the relay position
$z_r \in \mathcal{C}$ such that $z_r \in C^s_{T_1}$ and $\mathcal{Z}
\in C_{\cup}$.

Although Program (C) is reduced, it is still a non-convex
optimization problem. The objective function is non-convex and
different positions of the relay $z_r \in \mathcal{C}$ result in
different end node sets $T_1$ and $T_2$ for the hyperarcs of path
$\hat{l_1}$.

On the other hand, we know that the relay position is sensitive only
to the flow over path $\hat{l}_1$. In addition, as there always
exist a relay position $z_r \in  \mathcal{C}$ such that the min-cut
of path $\hat{l}_1$ is higher than that of path $\hat{l}_2$, then
this also holds true for $z^*_{\gamma \uparrow}$. Therefore,
optimizing the relay position to maximize the flow over path
$\hat{l}_1$ results in global optimal relay position solving the
original problem $(s,T,\mathcal{Z},\gamma)$. This motivates the
decoupling of computation of optimal relay position from the flow
maximization over the path $\hat{l}_1$.

\begin{proposition}\label{P2}
For a given problem instance $(s,T,\mathcal{Z},\gamma)$, if $g(\nu)
h(D_{sp^*})=D_{p^*}$, then $z^*_{\gamma \uparrow}=z_{p^*}$.
\end{proposition}
\vspace{1mm} Refer Appendix~\ref{ap:Proof-P2} for the detailed
proof. At point $p^{*}$, in general the following holds
$\frac{g(\mu)}{h(\pi^{p^*}_s)} \geq \frac{g(\nu)}{h(\pi^{p^*}_r)}$
(from Equation~(\ref{Sec2eq10})), thus making it naturally a good
candidate for $z^*_{\gamma \uparrow}$. Proposition~\ref{P2},
essentially proves that if the relay is positioned at $p^*$ and we
get $\frac{g(\mu)}{h(\pi^{p^*}_s)} = \frac{g(\nu)}{h(\pi^{p^*}_r)}$,
and if maximizing the flow over the path $\hat{l}_1$ results in no
spare source power (i.e. $g(\nu) h(D_{sp^*})=D_{p^*}$), then
$z^*_{\gamma \uparrow}=z_{p^*}$ and
$F^*=\frac{g(\mu)}{h(\pi^{p^*}_s)}$. Furthermore, the joint problem
in Program (C) can be reduced to solving in sequence the computation
of the optimal relay position $p^*$ by solving
Equation~(\ref{Sec2eq10}) and then calculating the max-flow $F^*$.
But this is not true when $\frac{g(\mu)}{h(\pi^{p^*}_s)} >
\frac{g(\nu)}{h(\pi^{p^*}_r)}$. We cover this case in the section of
algorithms.

Let us now see the problem in a different way. Consider the radius
$\pi_s \in (0,D_{st_n})$ and construct the hyperarc $C^s_{\pi_s}$.
Denote by $T'=\{t_{j} \in T|D_{st_{j}}
> \pi_s\}$, the set of destination nodes that
lie outside the hyperarc circle $C^s_{\pi_s}$. Then compute the
point $\widehat{r}$ such that \vspace{-2mm}
\begin{equation*}
z_{\widehat{r}}=\displaystyle\argmin_{z_{p} \in C^{s}_{\pi_s}}
(\displaystyle\max_{t_j \in T'} (D_{r' t_j})), \vspace{-2mm}
\end{equation*}
and position the relay at $\widehat{r}$ (here $\widehat{r}$ is the
point in $C^s_{\pi_s}$ such that the maximum among the distances to
the nodes in the set $T'$ from $\widehat{r}$ is minimized). If
$D_{s\widehat{r}} < \pi_s$, then we contract the hyperarc
$C^s_{\pi_s}$ to $C^s_{\widehat{r}}$, else we simply re-denote it
with $C^s_{\widehat{r}}$. Finally, we can construct the hyperarc
$C^{\widehat{r}}_{t_n}$ ( note that $\mathcal{Z} \in
C'_{\cup}=C^s_{\widehat{r}} \cup C^{\widehat{r}}_{t_n}$). The set
$\mathcal{R}'$ of points $\widehat{r}$ computed in this way for
different values of $\pi_s \in (0,D_{st_n})$ are the optimal relay
positions $z^{*}_{\gamma \uparrow}$ solving
$(s,T,\mathcal{Z},\gamma)$ for some $\gamma \in (0,\infty)$.
Figure~\ref{fig:Fig2}(a) captures this interesting insight of the
relationship between the points $\widehat{r}$ and $z^{*}_{\gamma
\uparrow }$. Note that the set $\widehat{\mathcal{R}}$ of points
$\widehat{r}$ is a discontinuous piecewise linear segment.

\subsection{Min-cost Problem $(s,T,\mathcal{Z},\gamma,F)$ And Duality}\label{sec:Mincost}
The min-cost problem $(s,T,\mathcal{Z},\gamma,F)$ can be written as
\begin{equation*}
\hspace{2mm}{\text{Minimize}}\hspace{3mm} (P^{s}_{T_1}+P^{r}_{T_2}+P^{s}_{T})  \hspace{35mm} \text{(D)}\\
 \end{equation*}
\vspace{-4mm}
\begin{align}
&\text{subject to:} \hspace{1mm} F \leq \min
\bigg(\frac{g({P^{s}_{T_1}})}{h(\pi_s)},\frac{g({P{r}_{T_2}})}{h(\pi_r)}
\bigg)+\frac{g({P^{s}_{T}})}{h(D_{st_n})}, \label{Sec3Beq1}\\
&\hspace{17mm} P^{s}_{T_1}+P^{s}_{T} \leq \mu, \hspace{2mm}
P^{r}_{T_2} \leq \nu. \label{Sec3Beq2}
\end{align}
In the non-convex Program (D), the path
$\hat{l}_1=\{C^s_{T_1},C^r_{T_2}\}$ correspond to the relay position
$z_r \in \mathcal{C}$ which is implicitly represented in the
distance variables $\pi_s$ and $\pi_r$. From Theorem~\ref{T1}, we
know that paths $\hat{l}_1$ and $\hat{l}_2$ carry all the min-cost
target multicast flow $F$. In this sub-section we refer the path
$\hat{l}_1$ as the cheapest path for a unit flow among all the paths
through $r$ in $L$ for given position of relay.

Now, we claim that $z^{*}_{F \downarrow } \in
\widehat{\mathcal{R}}$. This is true because given the hyperarc
$C^s_{T_1}$ of path $\hat{l}_1$ with optimal radius $\pi_s^*$, the
second hyperarc $C^r_{T_2}$ must be centered at the point that
minimizes the maximum among the distances to all the destination
nodes not spanned by the hyperarc $C^s_{T_1}$ from itself, as this
minimizes the cost over the hyperarc $C^r_{T_2}$. Therefore,
$z^{*}_{F \downarrow}$ (like $z^{*}_{\gamma \uparrow}$) always lie
on on the curve $\widehat{\mathcal{R}}$. This observation motivates
an interesting fundamental relationship between $z^{*}_{F \downarrow
}$ and $z^{*}_{\gamma \uparrow }$.

\begin{theorem}[Max-flow/Min-cost Duality] \label{T2}
For $F \in [0,F^{*}]$,
\begin{equation}\label{Sec3eq3}
z^*_{F \downarrow} = z^*_{\widehat{\gamma} \uparrow},
\end{equation}
where $\widehat{\gamma} \in
[\min(\overline{\gamma},\gamma),\max(\overline{\gamma},\gamma)]$ and
$z^*_{1 \downarrow} = z^*_{\overline{\gamma} \uparrow}$.
\end{theorem}
\vspace{2mm}

Theorem~\ref{T2} establishes the underlying duality relation between
the max-flow problem $(s,T,\mathcal{Z},\gamma)$ and the min-cost
problem $(s,T,\mathcal{Z},\gamma,F)$ and says that the point $z^*_{F
\downarrow}$ (or $z^*_{\widehat{\gamma} \uparrow}$) lies on the
segment $z^*_{1 \downarrow}-z^*_{F^* \downarrow}$
($z^*_{\overline{\gamma} \uparrow}-z^*_{\gamma \uparrow}$,
respectively) of the curve $\widehat{\mathcal{R}}$. Implying that
the optimal relay position $z^*_{F \downarrow}$ solving the problem
$(s,T,\mathcal{Z},\gamma,F)$ is also the optimal relay position
$z^*_{\widehat{\gamma} \uparrow}$ solving the problem
$(s,T,\mathcal{Z},\gamma)$ for some $\widehat{\gamma}$. The proof of
Theorem~\ref{T2} is presented in Appendix~\ref{ap:Proof-T2}.

However, the max-flow is not always reducible to a sequence of
decoupled problems. This is mainly due to the fact that the path
$\hat{l}_2$ can be cheaper than path $\hat{l}_1$ for a unit flow
corresponding to the optimal position $z^*_{F \downarrow}$, i.e.
\begin{equation*}
 g^{-1}(h(\pi^*_s))+g^{-1}(h(\pi^*_r)) \geq g^{-1}(h(D_{st_n})).
\end{equation*}
This information is not easy to get a priori. In contrast, we can
safely assume that
\begin{equation}\label{Sec3Beq4}
 g^{-1}(h(\pi^*_s))+g^{-1}(h(\pi^*_r)) \leq g^{-1}(h(D_{st_n})),
\end{equation}
as almost all wireless network models that comply with our model
result in the hyperarc cost function $g^{-1}(h(D_{uv_{k_u}}))$ being
the increasing convex function of distance $D_{uv_{k_u}}$ that
satisfy Inequality~(\ref{Sec3Beq4}). If Inequality~(\ref{Sec3Beq4})
holds, then similar to the Max-flow problem the joint optimal relay
positioning and min-cost flow optimization problem in Program (D)
can be reduced to a sequence of decoupled problems of computing the
optimal relay position and then optimizing the hyperarc powers to
achieve the min-cost flow $F$ in the network using the similar
arguments as in previous subsection. For a special of the min-cost
problem $(s,T,\mathcal{Z}, \gamma, F)$, we present the Min-cost
Algorithm that sequentially solves and outputs the optimal relay
position and powers to achieve the target flow $F \in [0,F^*]$ in
Section~\ref{sec:Mincostalg}.

%------------------------------------------------------------------------%
\section{ALGORITHMS} \label{sec:Algorithms}
%------------------------------------------------------------------------%

In this section we present the general max-flow and the special case
min-cost algorithms that solve the sequence of decoupled problems.
\subsection{Max-flow Algorithm}\label{sec:Maxflowalg}
\vspace{-4mm}
\begin{figure}[!h]
        \hrule height 0.75pt\vskip1pt
        \renewcommand{\algorithmicrequire}{\textbf{Input:}}
         \begin{algorithmic}[1]
        \REQUIRE Problem instance $(s,T,\mathcal{Z},\gamma)$.
        \vskip1pt\hrule\vskip2pt

\STATE Compute $p^{*}$, if $g(\nu)h(D_{sp^*})=g(\mu)h(D_{p^*t_n})$,
output $z^{*}_{\gamma \uparrow}=z_{p^*}$, $F^{*}=g(\nu)h(D_{sp^*})$
and quit, else go to $2$.

\STATE Construct the set $T'=\{t'_j \in T| D_{st'_j} < D_{p^*
t'_j}\}=\{t'_{1},..,t'_{j'}\}$ (ordered in increasing distance from
$s$) and compute $p^*_{T \backslash T'}$. If $D_{st'_{j'}} \leq
D_{sp^*_{T \backslash T'}}$, declare $z^{*}_{\gamma
\uparrow}=z_{p^*_{T \backslash T'}}$ and $F^{*}=g(\nu)h(D_{sp^*_{T
\backslash T'}})$ and quit, else go to Step $3$.

\STATE Compute the points $z^*_{1}$ and $z^*_{2}$, and maximized
multicast flow $F^*_{1}$ and $F^*_{2}$, respectively. Declare before
quitting,
\begin{equation*}
z^*_{\gamma \uparrow}=\begin{cases} z^*_{1} & \text{if $F^*_{1}
> F^*_{2}$,}\\
z^*_{2} & \text{if $F^*_{1} < F^*_{2}$.}
\end{cases}
\end{equation*}

        \vskip3pt\hrule\vskip3pt
        \renewcommand{\algorithmicensure}{\textbf{Output:}}
        \ENSURE $z^{*}_{\gamma \uparrow}$ and $F^{*}$.
         \end{algorithmic}
         \hrule height 0.75pt\vskip5pt
        \caption{Max-flow Algorithm.}
        \label{alg:Maxflowalg}
%        \vskip-5pt
\end{figure}
\vspace{-2mm} The Max-flow Algorithm in Figure~\ref{alg:Maxflowalg},
is a simple and non-iterative $3$ step algorithm that outputs the
optimal relay position and the maximized multicast flow. The first
step is essentially Proposition~\ref{P2}, in case it is not
satisfied the second step filters the redundant nodes that are too
close to the source and can be ignored. If the conditions of first
or second step are not met, then the third step divides the
computation of $z^*_{\gamma \uparrow}$ into two regions of
$\mathcal{C}$ and computes the optimal relay position $z^*_{1}$ and
$z^*_{2}$ for these two regions and outputs the better one. The
proof of optimality is provided in  Appendix~\ref{ap:Proof-A1}.

\subsection{Min-cost Algorithm}\label{sec:Mincostalg}
In this subsection, we assume that the Inequality~(\ref{Sec3Beq4})
is satisfied and the target flow $F \in [0,F^*]$ goes over the path
$\hat{l}_1$ (corresponding to the optimal relay position $z^*_{F
\downarrow}$) only. Min-cost Algorithm in
Figure~\ref{alg:Mincostalg}, unlike the Max-flow algorithm, is an
iterative algorithm. In the first step the geometric feasibility
region is constructed and in the second step this region is divided
into at most $n-1$ sub-regions. The optimal relay position is
computed for all the sub-regions and the one minimizing the cost
among them is declared global optimal. Computing the optimal relay
position for the sub-regions is a simple geometric convex program
that can be solved efficiently and the number of such iterations is
upper bounded by $n-1$. The proof of optimality is presented in
Appendix~\ref{ap:Proof-A2}.
\begin{figure}[!ht]
        \hrule height 0.75pt\vskip1pt
        \renewcommand{\algorithmicrequire}{\textbf{Input:}}
         \begin{algorithmic}[1]
        \REQUIRE Problem instance $(s,T,\mathcal{Z},\gamma,F)$ and $C'_{\cap}$.
        \vskip1pt\hrule\vskip2pt

\STATE Compute $\widehat{p} = \displaystyle\argmin_{p \in C'_{\cap}}
(h(D_{sp}) + \displaystyle\max_{i \in [1,n]}(h(D_{pt_i})))$, and
build the set $\widehat{T}=\{\widehat{t} \in T| D_{s{\widehat{t}}}
\leq D_{s{\widehat{p}}}\}$. If $\widehat{T}\neq \{\emptyset\}$, then
recompute $\widehat{p} = \displaystyle\argmin_{p \in C'_{\cap}}
(h(D_{sp}) + \displaystyle\max_{t \in T\backslash
\{\widehat{T}\}}(h(D_{pt})))$, calculate
$\Psi_{\widehat{p}}=h(D_{s\widehat{p}})+D_{\widehat{p} t_n}$ and to
go to Step $2$.

 \STATE Build the set $\overline{T}=\{t \in T \backslash
\{\widehat{T},t_n\}|D_{st}>\pi^{\widehat{p}}_s, D_{st} \leq
\pi'_s\}=\{\overline{t}_1,.., \overline{t}_l\}$ (ordered in
increasing distance from
 $s$), compute the points
\begin{equation*}
\widehat{p}_j = \displaystyle\argmin_{p \in \overline{C}^s_{j}}
(\max(h(D_{sp}),h(D_{s\overline{t}_{j-1}})) + \displaystyle\max_{t
\in \overline{T}_j}(h(D_{pt}))), \vspace{-2mm}
\end{equation*}
and calculate the cost of unit flow
$\overline{\Psi}_j=h(D_{s\widehat{p}})+ h(\displaystyle\max_{t \in
\overline{T}_j}(D_{\widehat{p}t}))$ over the path $\hat{l}_2$
corresponding to the relay position $\widehat{p}_j$,  $\forall j \in
[1,l]$. Declare
\begin{equation*}
z^*_{F \downarrow}=
\begin{cases}
z_{\widehat{p}} & \text{if $\Psi_{\widehat{p}} \leq \overline{\Psi}_m$,}\\
z_{\overline{p}_m} & \text{if $\Psi_{\widehat{p}} \geq
\overline{\Psi}_m$},
\end{cases}
\end{equation*}
where $\overline{\Psi}_m=\displaystyle\min_{j \in
[1,n]}(\overline{\Psi}_j)$, ${P^{s}_{T_1}}^*=g^{-1}(h(\pi^*_{s})F)$
and ${P^r_{T_2}}^*=g^{-1}(h(\pi^*_{r})F)$ and quit.

        \vskip3pt\hrule\vskip3pt
        \renewcommand{\algorithmicensure}{\textbf{Output:}}
        \ENSURE $z^{*}_{F \downarrow}$, ${P^{s}_{T_1}}^*$ and ${P^{r}_{T_2}}^*$.
         \end{algorithmic}
         \hrule height 0.75pt\vskip5pt
        \caption{Min-Cost Algorithm.}
        \label{alg:Mincostalg}
%        \vskip-5pt
\end{figure}

%------------------------------------------------------------------------%
\section{Example: Low-SNR Achievable Network Model} \label{sec:Example}
%------------------------------------------------------------------------%
In this section we present an example from the interference
delimited network model that was originally presented in
\cite{Thakur-Fawaz-Medard-arXivISIT2011}.
\subsection{Low-SNR Broadcast and MAC Channel Model} Consider the
AWGN Low-SNR (wideband) Broadcast Channel with a single source $s$
and multiple destinations $T=\{t_1,..,t_n\}$ (arranged in the order
of increasing distance from $s$). From \cite{Cover-1972} and
\cite{ElGamal-Cover-1980}, we know that the superposition coding is
equivalent to time sharing, which is optimal. Implying that the
broadcast communication from a single source to multiple receivers
can be decomposed into communication over $n$ hyperarcs sharing the
common source power. Therefore, we get the set of hyperarcs
$\mathcal{A}_{bc}=\{(s,t_1),(s,t_1t_2),..,(s,t_1t_2..t_n)\}$.

Similarly, in the Low-SNR (wideband) regime, interference becomes
negligible with respect to noise, and all sources can achieve their
point-to-point capacities analogous to Frequency Division Multiple
Access (FDMA). In general, the MAC Channel consisting from $n$
sources $s_1,...,s_n$ transmitting to a common destination $t$ can
be interpreted as $n$ point-to-point arcs each having point-to-point
capacities. Thus, we get $\mathcal{A}_{mac}=\{(s_1,t),..,(s_n,t)\}$.
Each hyperarc $(s,t_1..t_j) \in \mathcal{A}_{bc} \cup
\mathcal{A}_{mac}$ is associated with the rate function
\begin{equation}\label{Exeq1}
R^s_{t_j}= \frac{P^s_{t_j}}{N_{0}D^{\alpha}_{st_j}}, \forall j \in
[1,n],
\end{equation}
where $\alpha \geq 2$ is the path loss exponent.

\subsection{Low-SNR Achievable Hypergraph Model}
By concatenating the Low-SNR Broadcast Channel and MAC Channel
models we obtain an Achievable Hypergraph Broadcast Model. For
example the Broadcast Relay Channel consisting of a single source,
$n$ destinations and a relay. Although, the time sharing and FDMA
are capacity achieving optimal schemes in the respective models, the
Achievable Hypergraph Model is not necessarily capacity achieving.
In contrast and more importantly for practical use, this model is
easy to scale to larger and more complex networks.

The above Low-SNR Achievable Hypergraph Model also incorporates
fading \cite{Thakur-Fawaz-Medard-arXivISIT2011}. The rate function
in Equation~(\ref{Exeq1}) is linear in transmitter power and convex
in hyperarc distance, hence the results from this paper can be
directly applied.

%------------------------------------------------------------------------%
\section{CONCLUSION} \label{sec:Conclusion}
%------------------------------------------------------------------------%
We present simple and efficient geometry based algorithms for
solving joint relay positioning and flow (max-flow/min-cost)
optimization problems for a fairly general class of hypergraphs. Any
application that satisfies the hypergraph construction rules and can
be modeled under the classical multicommodity framework can use the
results presented here.

As a part of future work it would be of interest to extend the work
presented here to the general multicommodity setting where multiple
sessions use a common relay.

 \bibliographystyle{./../../Biblio/IEEEtran}

 \bibliography{./../../Biblio/IEEEabrv,./../../Biblio/bibLowSNR,./../../Biblio/bibNC,./../../Biblio/bibCellularStd,./../../Biblio/bibOptim}

\clearpage
\appendices

% %------------------------------------------------------------------------%
% \section{Notations}
% %------------------------------------------------------------------------%
% The max-flow problem instance is given by $(s,T,\gamma)$, where the
% source node $s$, the destination node set $T=\{t_1,..,t_n\}$ are
% placed on a $2$-D Euclidean plane. The total source and relay node
% powers are given as $\mu$ and $\nu$, respectively, where
% $\gamma=\frac{\nu}{\mu}$. $\mathcal{C}$ denotes the convex hull of
% the nodes $\{s,T\}$. For a given position of relay node $r \in
% \mathcal{C}$, let the network hypergraph is denoted by
% $\mathcal{G(N,A)}$. The set of paths from $s$ to a destination
% $t_{i} \in T$ is given by
% $L^{z_r}_{i}=\{l^{z_r}_{i,1},..,l^{z_r}_{i,K_i}\}$. Each
% $l^{z_r}_{i,k_i} \in L^{z_r}_{i}$ from $s$ to $t_i$ with $k_{i} \in
% [1,K_{i}]$, either goes through $r$ and
% $l^{z_r}_{i,k_{i}}=\{(s,J^{s}_{i,k_{i}}),(r,J^{r}_{i,k_{i}})\}$,
% where $t_i \in J^{r}_{i,k_i}$, or is a single hyperarc path
% $l^{z_r}_{i,k_{i}}=\{(s,J^{s}_{i,k_{i}})$, where $t_i \in
% J^{s}_{i,k_i}$. The set of paths from $s$ that spans all the
% destination set $T$ is denoted by
% $L^{z_r}=\{l^{z_r}_{1},..,l^{z_r}_{K}\}$, where all the paths go
% through $r$ except for one path i.e. the biggest hyperarc emanating
% from $s$ that spans all the destination set.

%------------------------------------------------------------------------%
\section{Proof of Proposition~\ref{P1} }\label{ap:Proof-P1}
%------------------------------------------------------------------------%
\begin{IEEEproof}
Let the set of nodes $\mathcal{N}=\{s,r,t_1,..,t_n\}$ be placed on
the $2$-D Euclidean plane and $\mathcal{C}$ denote their convex hull
polygon. Let us assume that the relay node $r$ is placed outside the
polygon $\mathcal{C}$, i.e. $z_r \notin \mathcal{C}$ and $c$ be the
nearest point to $r$ in the polygon $\mathcal{C}$. Let the line
segment joining $z_r$ and $c$ be denoted as $z_r - c$.

The rate over all the hyperarcs that either emanate from $r$ or $r$
is the farthest end node of the hyperarc, is relay position
dependent. As the hyperarc rate is a decreasing function of
distance, moving the relay closer to $c$ on the segment $z_r - c$
decreases the distance between $r$ and every point in the polygon
$\mathcal{C}$ and thus to every node in the system. This implies
that for a given power allocation for the relay position dependent
hyperarcs the rate can be increased as the relay gets closer to the
point $c$. Consequently, we can conclude that the optimal relay
position $z^{*}_{\gamma \uparrow }$ maximizing the multicast flow
$F$ for the session $(s,T)$ will lie in the convex hull polygon
$\mathcal{C}$.

Similarly, all the relay position dependent hyperarcs will need
lesser power to carry a given flow of value $F$ as the relay $r$
moves closer on the line segment $z_r -c$ to the the point $c$.
Implying, that for any target flow $F$ the optimal relay position
$z^{*}_{F \downarrow }$ will lie in $\mathcal{C}$. This concludes the proof.
\end{IEEEproof}

%------------------------------------------------------------------------%
\section{Proof of Theorem~\ref{T1} }\label{ap:Proof-T1}
%------------------------------------------------------------------------%

Before we formally prove Theorem~\ref{T1}, we need to establish some
basic tools from convex analysis.

Let $f: \mathbf{R}^+ \longrightarrow \mathbf{R}^+$ be an increasing
and convex function that maps a non-negative real input to a
non-negative real output. Denote with $\bar{f}(x)$ the
sub-derivative of $f(x)$ at the point $x \in \mathbf{R}$ and let
$\partial f(x)$ denote the complete set of sub-derivatives at point
$x$. If the set $\partial{f(x)}$ is a singleton set, then
$\bar{f}(x)=\frac{\partial{f}}{\partial{x}}$, which is simply the
derivative of the $f$ at $x$; else there exist a finite interval
$\partial f(x)$ between the left and right limits of $f$ at $x$. In
addition, let us also assume that $f(0) \leq 0$. Then the following
proposition is true.
\begin{proposition}\label{T1P1}
If $f$ is any increasing convex function such that $f(0) \leq 0$
then
\begin{equation}
\displaystyle\sum_{i=1}^{n} f(x_{i}) \leq f
\left(\displaystyle\sum_{i=1}^{n} x_{i} \right),
\hspace{2mm} (x_{1},..,x_n) \in \mathbf{R}^{n+}.
\end{equation}
\end{proposition}
\begin{IEEEproof}
As $f$ is increasing over the real line, for $x_1 < x_2$ we have
$f(x_1) \leq f(x_2)$. Also, as $f$ is convex $\bar{f}(x_1) \leq
\bar{f}(x_2)$.

Let the slopes of line joining the points $(0,f(0))$ and $(x_1,f(x_1))$,
$(x_1,f(x_1))$ and $(x_2,f(x_2))$ be given by,
\begin{equation}\label{T1eq1}
s_{1} = \frac{f(x_1)-f(0)}{x_1}, \hspace{2mm} s_{2} = \frac{f(x_2)-f(x_1)}{x_2 -x_1},
\end{equation}
where $0<x_1<x_2$ are points on real line. From the Generalized Mean
Value Theorem we know that there always exist a point $c$ and $c'$
between $[0,x_1]$ and $[x_1,x_2]$, such that $\bar{f}(c)=s_1$ and
$\bar{f}(c')=s_2$, respectively. This, along with the fact that $f$
is increasing and convex implies,
\begin{equation*}
\bar{f}(0) \leq \bar{f}(c)=s_1 \leq \bar{f}(x_1), \hspace{2mm}
\bar{f}(x_1) \leq \bar{f}(c')=s_2 \leq \bar{f}(c').
\end{equation*}
In general, given $n$ points $x_1 < ... < x_n$ with $s_{i}$ as the
slope of line joining the points
$(x_{i},f(x_{i}))-(x_{i+1},f(x_{i+1}))$ we get,
\begin{equation}
s_1 \leq s_2 \leq ... \leq s_{n-1}. \label{T1eq2}
\end{equation}
Consider now the four points $(0,f(0))$, $(x_1,f(x_{1}))$,
$(x_{2},f(x_{2}))$ and $(x_{12}f(x_{12}))$, where
$x_{12}=x_{1}+x_{2}$. From Inequality~\ref{T1eq2} we get,
\begin{equation}\label{T1eq3}
\frac{f(x_{1})-f(0)}{x_{1}} \leq \frac{f(x_{2})-f(x_{1})}{x_{2}-x_1}
\leq \frac{f(x_{12})-f(x_{2})}{x_{1}}.
\end{equation}
Inequality~\ref{T1eq3} implies,
\begin{equation}
f(x_1)+f(x_2)-f(0) \leq f(x_{12})=f(x_{1}+x_{2}). \label{T1eq4}
\end{equation}
If $f(0) = 0$, then $f(x_1)+f(x_2) \leq f(x_1 + x_2)$. Using this
fact it is straightforward to show that this also holds for $f(0)
<0$, for any $(x_1,x_2) \in \mathbf{R}^{2+}$.

Without loss of generality, assume that $x_{12} < x_3$,
repeating the previous step of Inequalities~\ref{T1eq3}-\ref{T1eq4} we
get,
\begin{equation*}
\begin{split}
f(x_{12})+f(x_3)& \leq f(x_{123} ) \Rightarrow \\
f(x_{1})+f(x_{2})+f(x_3) & \leq f(x_{1}+x_2 +x_3),
\end{split}
\end{equation*}
where $x_{123}=x_1+x_2+x_3$. Similarly, repeating these $n$ times we have
\begin{equation}
\displaystyle\sum_{i=1}^{n} f(x_i) \leq f
\left(\displaystyle\sum_{i=1}^{n} x_i \right), \hspace{2mm} (x_{1},..,x_n) \in \mathbf{R}^{n+},\label{{T1eq5}}
\end{equation}
if $f(0) \leq 0$. This proves the proposition.
\end{IEEEproof}

\vspace{1mm}

\begin{figure}[tp]
\begin{center}
\psfrag{a}{{$(a)$}} \psfrag{b}{{$(b)$}} \psfrag{x}{\tiny{$x$}}
\psfrag{x'}{\tiny{$x'$}} \psfrag{fx}{\tiny{$f(x)$}}
\psfrag{x1}{\tiny{$x_1$}} \psfrag{x2}{\tiny{$x_{2}$}}
\psfrag{x3}{\tiny{$x_{3}$}} \psfrag{f1}{\tiny{$f(x_1)$}}
\psfrag{f2}{\tiny{$f(x_{2})$}} \psfrag{f3}{\tiny{$f(x_{3})$}}
\includegraphics[width=1\columnwidth]{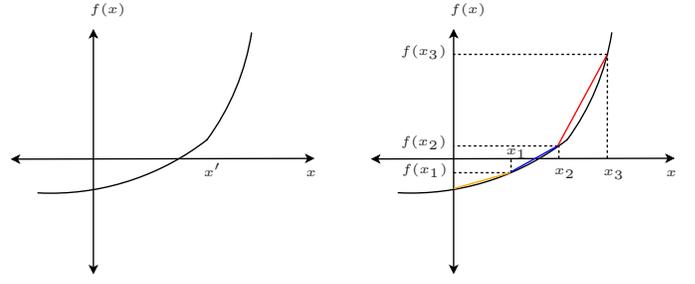}
\end{center}
\vspace{-4mm} \caption{{(a): Increasing convex function $f$
(possibly non-differentiable, e.g. at $x'$). (b): Orange, blue and
red lines joining the points $(f(0),0)-(f(x_1),x_1)$,
$(f(x_1),x_1)-(f(x_2),x_2)$ and $(f(x_3),x_3)-(f(x_2),x_2)$,
respectively.}} \label{fig:Fig3} \vspace{-6mm}
\end{figure}

Now let $f$ and $g$ be increasing convex functions satisfying Proposition~\ref{T1P1} and define $f_{i}(x_i)=\lambda_{i} f(x_i)$ and
$g_{i}(y_i)=\lambda'_{i} g(y_i)$ as the $2n$ linear compositions of
the function $f$ and $g$ for $i \in [1,n]$, where $\lambda_{i} \geq 0$,
$\lambda'_{i} \geq 0$, $\forall i \in [1,n]$ and $f(0)=g(0)=0$. Then
consider the following program,
\begin{align*}
\hspace{10mm} \mbox{Maximize}& \hspace{5mm} \left(
\mathcal{F}_n=\displaystyle\sum_{i=1}^{n}
\min(f_{i}(x_{i}),g_{i}(y_{i})) \right) \hspace{1mm} \mbox{(T1A)}\\
\mbox{subject to:}& \hspace{2mm}
\displaystyle\sum_{i=1}^{n} x_{i} \leq \mu, \hspace{2mm} \displaystyle\sum_{i=1}^{n} y_{i} \leq \nu, \hspace{25mm}\\
& \hspace{2mm} x_{i} \geq 0, \hspace{2mm} y_{i} \geq 0, \hspace{5mm} \forall i \in [1,n]. \hspace{10mm}\\
\end{align*}
Denote the set $S^*$ as the set of optimizers of Program (T1A). In
addition, assume that
\begin{equation}
\min(\lambda_{k},\lambda'_{k}) = \displaystyle\max_{i \in [1,n]}
\left( \min(\lambda_{i},\lambda'_{i}) \right), \hspace{2mm}  k \in
[1,n].\label{T1eq6}
\end{equation}
Let us denote a set of points
$U^*=\{(\mathbf{x}_k,\mathbf{y}_k)=(0,..,0,\mu,0,..,0),(0,..,0,\nu,0,..,0)\}$
where, $\mathbf{x}_k$ and $\mathbf{y}_k$ are the vectors with $\mu$
and $\nu$ at the $k^{th}$ place and all other elements $0$,
respectively. Then the following proposition is true.
\begin{proposition}\label{T1P2}
$U^{*} \subset S^{*}.$
\end{proposition}
\begin{IEEEproof}
Consider program (T1A) and without loss of generality assume that
\begin{align}
\min(\lambda_{1},\lambda'_{1}) \leq \min(\lambda_{2},\lambda'_{2})
\leq .... \leq \min(\lambda_{n},\lambda'_{n}), \label{T1eq7}
\end{align}
and
\begin{align}
\min(\lambda_{k},\lambda'_{k}) =
\min(\lambda_{k+1},\lambda'_{k+1})=..
=\min(\lambda_{n},\lambda'_{n}), \label{T1eq8}
\end{align}
where Equation~(\ref{T1eq8}) says that the last $n-k$ terms of
Inequality~(\ref{T1eq7}) are equal. Let
\begin{equation*}
\rho_{i} =\begin{cases}
            f_{i}(x_{i})=\lambda_i f(x_i) & \text{if $\lambda_{i} \leq \lambda'_{i}$,}\\
            g_{i}(y_{i})=\lambda'_i g(y_i) & \text{if $\lambda_{i} > \lambda'_{i}$.}
            \end{cases}
\end{equation*}
Then, from Proposition~\ref{T1P1} for any $0 \preceq (\epsilon,\epsilon) \in \mathbf{dom}(f,g)$ we get,
\begin{equation*}
\displaystyle\sum_{i =1}^{n} \rho_i (\epsilon) \leq \rho_j (n
\epsilon),
\end{equation*}
which implies
\begin{equation}
\displaystyle\sum_{i =1}^{n} \min(f_i(\epsilon),g_i(\epsilon)) \leq
\min(f_j(n \epsilon),g_j(n \epsilon)), \label{T1eq9}
\end{equation}
as $\rho_{i}$ is the limiting sub-term of the $i^{th}$ term
$\min(f_{i}(x_{i}),g_{i}(y_{i}))$ of function $\mathcal{F}_n$ for
any $i \in [1,n]$ and $j \in [k,n]$. From Inequality~(\ref{T1eq9}), we
can infer that simply maximizing $\min(f_j(x_j),g_j(y_j))$ (for any $j
\in [k,n]$) alone maximizes $\mathcal{F}_n$ in Program (A), with all
other terms attaining the value $0$ (except for $j^{th}$ term)
because $\lambda_{i}f(0)=\lambda'_{i}f(0)=0, \forall i \in
[1,n]$. Therefore, $\mathcal{F}^{*}_{n}=\min(f_j(\mu),g_j(\nu))$ for any $j\in [k,n]$ is
the maximum value of the function $\mathcal{F}_n$ in Program (T1A).

Hence $U^* \subset S^*$, where
$U^*=\{(\mathbf{x}_j,\mathbf{y}_j)=(0,..,0,\mu,0,..,0),(0,..,0,\nu,0,..,0)|j
\in [k,n]\}$.
\end{IEEEproof}
\vspace{1mm}

Under the same setting as for Program (T1A), consider the following program,
\begin{align*}
\hspace{10mm} \mbox{Minimize}& \hspace{5mm} \left(
\mathcal{F}'_n=\displaystyle\sum_{i=1}^{n}
(x_{i}+y_{i}) \right) \hspace{17mm} \mbox{(T1B)}\\
\mbox{subject to:}& \hspace{2mm} \zeta \leq \displaystyle\sum_{i=1}^{n} \min(f_i(x_{i}),g_i(y_i)),  \hspace{25mm}\\
& \displaystyle\sum_{i=1}^{n} x_{i} \leq \mu, \hspace{2mm} \displaystyle\sum_{i=1}^{n} y_{i} \leq \nu, \\
& x_{i} \geq 0, \hspace{2mm} y_{i} \geq 0, \hspace{5mm} \forall i \in [1,n], \hspace{10mm}\\
\end{align*}
where $\zeta \geq 0$ is a given positive real number and such that
\begin{equation}\label{T1eq20A}
\zeta \leq \min(f_{i}(\mu),g_{i}(\nu)), \hspace{2mm} \forall i \in [1,n].
\end{equation}
Denote with $S'^*$, the set of optimizers of Program (T1B). Then the
following Proposition holds true.
\begin{proposition}\label{T1P3}
$\mathbf{(x^{\zeta}_{k'},y^{\zeta}_{k'})} \subset S'^{*}$.
\end{proposition}
\begin{IEEEproof}
% Let us assume that
% \begin{equation}\label{T1eq20}
% \lambda_{k'}+\lambda'_{k'} =
% \displaystyle\max_{i \in [1,n]} (\lambda_{i}+\lambda'_{i}).
% \end{equation}
Let us assume that the inverse
functions $f^{-1}_{i}$ and $g^{-1}_{i}$ exists such that $f^{-1}_{i}(\lambda_i
f(x_i))=x_i$ in addition to $g^{-1}_{i}(\lambda_i g(y_i))=y_i$, for all
$i \in [1,n]$. Also assume that for any $\epsilon \geq 0$
we get,
\begin{equation}\label{T1eq21}
\begin{split}
f^{-1}_{k'}(\lambda_{k'} f(x^{\epsilon}_{k'}))+g^{-1}(\lambda_{k'} g(y^{\epsilon}_{k'}))&=x^{\epsilon}_{k'}+y^{\epsilon}_{k'} \\
&= \displaystyle\min_{i \in [1,n]} (x^{\epsilon}_{i}+y^{\epsilon}_{i}),
\end{split}
\end{equation}
where
$\lambda_{i}f(x^{\epsilon}_i)=\lambda'_{i}g(x^{\epsilon}_i)=\epsilon,
\forall i \in [1,n]$. Since, $f$ and $g$ are increasing convex
functions for non-negative input, their inverses $f^{-1}$ and
$g^{-1}$ are increasing concave functions for non-negative input.
From Proposition~\ref{T1P1} we can deduce the reverse inequality for
concave functions,
\begin{equation}\label{T1eq21A}
\displaystyle\sum_{i=1}^{n} f'(x_i) \geq f'
\left(\displaystyle\sum_{i=1}^{n} x_i \right),
\end{equation}
where $f':\mathbf{R}^+ \longrightarrow \mathbf{R}^+$ is an increasing concave function. From Inequality~(\ref{T1eq21A}), we
get for any $\epsilon \geq 0$
\begin{equation}
\begin{split}\label{T1eq22}
f^{-1}(\lambda_{k'} f(&nx^{\epsilon}_{k'}))+g^{-1}(\lambda_{k'} g(ny^{\epsilon}_{k'})) \\
&\leq \displaystyle\sum_{i=1}^{n} \left(f^{-1}(\lambda_{i} f(x^{\epsilon}_{i}))+g^{-1}(\lambda_{i} g(y^{\epsilon}_{i})) \right).
\end{split} \end{equation}
Lastly, from Inequality~(\ref{T1eq20A}) and (\ref{T1eq22}) we conclude that  $(\mathbf{x^{\zeta}_{k'}},\mathbf{y^{\zeta}_{k'}}) \subset S'^{*}$,
where $\mathbf{x^{\zeta}_{k'}}=(0,..,0,\zeta,0,...,0)$ and
$\mathbf{y^{\zeta}_{k'}}=(0,..,0,\zeta,0,...,0)$.
\end{IEEEproof}
\vspace{1mm}

Note that we did not assume the differentiability of the functions
$f$ and $g$.

%\begin{figure}[tp]
%\begin{center}
%\psfrag{s}{$s$} \psfrag{r}{$r$} \psfrag{t1}{$t_1$}
%\psfrag{t2}{$t_{2}$} \psfrag{3}{\tiny{$t_{3}$}} \psfrag{a}{$(a)$}
%\psfrag{b}{$(b)$} \psfrag{c}{$(c)$} \psfrag{j1}{\tiny{$J_{1}$}}
%\psfrag{jr}{\tiny{$J_{r}$}} \psfrag{j2}{\tiny{$J_{2}$}}
%\psfrag{j3}{\tiny{$J_{3}$}} \psfrag{j'1}{\tiny{$J'_{1}$}}
%\psfrag{j'2}{\tiny{$J'_{2}$}} \psfrag{j'3}{\tiny{$J'_{3}$}}
%\includegraphics[width=1\columnwidth]{Pictures/T1.eps}
%\end{center}
%\vspace{-4mm} \caption{{Problem instance with $\{s,r,t_1,t_2\}$.
%(a): Source hyperarcs - $\{(s,r),(s,rt_1),(s,rt_1t_2)\}$. (b): Relay
%hyperarcs (dashed arrows) - $\{(r,t_2),(r,t_2t_1)\}$. (c):
%Hypergraph $\mathcal{G(N,A)}$.}} \label{fig:T1} \vspace{-6mm}
%\end{figure}

The set of paths $L$, that span all the destinations will be central
to the proof of Theorem~\ref{T1}, so let us clarify some more
notations. The path set $L=\{l_1,...,l_{\tau}\}$ contains only one
path from $s$ to $T$ that does not go through $r$, namely
$\{(s,t_1,..,r,..,t_n)=(s,T)\}$ and without loss of generality let
us assume $l_{\tau}=(s,T)$. All other paths go through $r$ and
consist of two hyperarcs, i.e.
$l_j=\{(s,T^{j}_{1}),(r,T^{j}_{2})\}$ where $r\in T^{j}_{1}$
and $T^{j}_{1} \cup T^{j}_{2}=T, \forall j \in [1,\tau-1]$. Then the
path flow is given by
\begin{equation*}
F_{l_j}=\begin{cases} \min(R^s_{T^{j}_{1}},R^r_{T^{j}_{2}})
&=\min(\lambda_{j}^{1}g(P^s_{T^{j}_{1}}),g(
\lambda_{j}^{2} P^r_{T^{j}_{2}})) \\
& \text{if $j \in [1,\tau-1]$,}\\
R^s_{T} =\lambda_{{j}}g(P^s_{T}) & \text{if $j =
\tau$,}
\end{cases}
\end{equation*}
where $\lambda_{j}^{1}=\frac{1}{h(D_{st^{j}_{1}})}$,
$\lambda_{j}^{2}=\frac{1}{h(D_{rt^{j}_{2}})}$, $\forall j \in
[1,\tau-1]$  and $\lambda_{{{\tau}}}=\frac{1}{h(D_{st_n})}$, where
$t^j_1 \in T$ and $t^j_2 \in T$ are the farthest nodes from $s$ and
$r$ spanned by the hypearcs $(s,T^j_1)$ and $(r,T^j_2)$,
respectively.

\vspace{1mm}
\begin{IEEEproof}[Proof of Theorem 1]
Consider the hypergraph $\mathcal{G(N,A)}$ for the given position of relay
$z_r \in \mathcal{C}$ and the path
based formulation of multicast max-flow and min-cost problems in
Program (A) and (B), respectively.

Since the hyperarcs are constructed in the order of increasing
distance from the transmitter, there exist no two paths from the
$s$ to any destination $t_i \in T$ that are edge disjoint. This
implies that only the paths spanning all the destinations in the set
$L$ should to be considered, as sending any information over the
paths that span a subset $T' \subset T$ has to be resent over at
least one path in the set $L$ that spans all the set $T$. This fact
reduces Programs (A) and (B) to
\begin{center}
    \begin{tabular}{ | l | l |}
    \hline
  {\scriptsize{Max-flow}} & {\scriptsize{Min-cost}} \\ \hline
    {\scriptsize{Maximize $F =\displaystyle\sum_{l\in L}
F_{l}$ \hspace{2mm} \mbox{(T1C)}}} & {\scriptsize{Minimize
\hspace{4mm} $P$ \hspace{11mm} \mbox{(T1D)}}}\\

{\scriptsize{subject to: $F_{i} \leq \displaystyle\sum_{l\in L}
F_{l}, \forall i \in [1,n]$,}} & {\scriptsize{subject to: $F_{i}
\leq
\displaystyle\sum_{l \in L} F_{l}, \forall i \in [1,n]$, }}\\

\hspace{11mm}{\scriptsize{$F_{l} \geq 0, \hspace{2mm} \forall l \in
L.$}}
 & \hspace{11mm}{\scriptsize{$F \leq F_{i}, \hspace{2mm} \forall i \in [1,n]$,}}\\
  & \hspace{11mm}{\scriptsize{$F_{l}  \geq 0, \hspace{2mm}\forall l \in L$, }}\\ \hline
    \end{tabular}
\end{center}
where $P=\sum_{j=1}^{\tau-1} (P^s_{T^{l_j}_{1}}+P^s_{T^{l_j}_{2}}) +
P^{s}_{T}$ in Program (T1C), i.e. the sum of powers of all the
hyperarcs of all the paths in $L$. Therefore,
\begin{equation*}
F_{i} =\sum_{l\in L} F_{l}, \hspace{2mm} \forall i
\in [1,n], \Longrightarrow F=\min_{i \in [1,n]}F_i=\sum_{l\in L} F_{l}.
\end{equation*}

Without loss of generality let us assume that among all the paths
from $s$ through $r$ to $T$ the path $l_{k} \in L$ has the highest
min-cut, i.e. $\min(\lambda_{l_k}^{1},\lambda_{l_k}^{2})\geq \max_{j
\in [1,\tau -1]} \lambda_{l_j}^{1},\lambda_{l_j}^{2}$. Then we get
two scenarios, either
\begin{equation}\label{T1eq23}
\min(\lambda_{1}^{1},\lambda_{1}^{2}) \leq .. \leq
\min(\lambda_{{k}}^{1},\lambda_{{k}}^{2}) \leq \lambda_{{\tau}},
\end{equation}
where, the last inequality of Inequalities~(\ref{T1eq23}) says that
the path $(s,T)$ has the highest min-cut among all the paths in $L$.
Then from Proposition~\ref{T1P2}, the multicast flow can be
maximized by simply maximizing the flow over the path
$l_{\tau}=(s,T)$, and since maximizing the flow over this path
consumes all the source power $\mu$ the optimal multicast flow
$F^{*}$ is given by
\begin{equation}\label{T1eq24}
F^*=F^{*}_{l_{\tau}}=R^{s*}_{T}=\lambda_{\tau} g(\mu).
\end{equation}
 Otherwise if,
\begin{equation}\label{T1eq25}
\min(\lambda_{1}^{1},\lambda_{1}^{2}) \leq ..\leq \lambda_{{\tau}}
\leq.. \leq \min(\lambda_{{k}}^{1},\lambda_{{k}}^{2}),
\end{equation}
then again by Proposition~\ref{T1P2} maximizing the flow only over path
$l_k$ maximizes the multicast flow $F$ in Program (T1C). Thus, we get
\begin{equation}\label{T1eq26}
F^{*}_{l_{k}}=\min(R^{s*}_{T^{k}_{1}},R^{r*}_{T^{k}_{2}})=\min(\lambda^1_{k}
g(\mu),\lambda^2_{k} g(\nu)).
\end{equation}
Furthermore, if $\lambda_{{k}}^{1}g(\mu) < \lambda_{{k}}^{2}g(\nu)$,
i.e. if the source has relatively more power than relay $r$, then
the rest of the flow must be send over the path $(s,T)$ as any other
path through the relay (i.e. $l_{j}$ where $j \neq k$ and $j\in
[1,\tau -1]$) cannot be used due to no spare power left with relay.
This implies
\begin{equation}\label{T1eq27}
\begin{split}
F^{*}&=\min(R^{s*}_{T^{k}_{1}},R^{r*}_{T^{k}_{2}}) + \hat{R^{s}_{T}}\\
&=\min(\lambda^1_{k} g(\mu'),\lambda^2_{k} g(\nu))+ \lambda_{\tau}
g(\mu-\mu'),
\end{split}
\end{equation}
where $\lambda_{{k}}^{1}g(\mu') = \lambda_{{k}}^{2}g(\nu)$ and
$\hat{R^{s}_{T}}=R^{s}_{T} (P^{s}_{T}=\mu-\mu')=\lambda_{\tau}
g(\mu-\mu')$. Thus, all the maximized multicast flow $F^*$ goes over
at most two paths, $l_{k}$ and $l_{\tau}$. Integrating
Equations~(\ref{T1eq26}) and (\ref{T1eq27}), we get
\begin{equation}\label{T1eq28}
F^{*}=
\begin{cases}
R^{s*}_{T} \hspace{5mm} \text{if $\lambda_{\tau}= \displaystyle\max_{j\in[1,\tau]} (\min(\lambda^{1}_{{j}},\lambda^{2}_{{j}}))$,}\\
\min(R^{s*}_{T^{k}_{1}},R^{r*}_{T^{k}_{2}}) + \hat{R^{s}_{T}} \\
\hspace{5mm}\text{if $\min(\lambda^{1}_{{k}},\lambda^{2}_{{k}}) =
\displaystyle\max_{j \in [1,\tau]}
(\min(\lambda^{1}_{{j}},\lambda^{2}_{{j}})).$}
\end{cases}
\end{equation}
From Equation~(\ref{T1eq28}), we conclude that for any given
position of relay $z_r \in \mathcal{C}$, the optimal multicast
max-flow $F^*$ goes over at most two paths namely $l_{k}$ and
$l_{\tau}$. Consequently, this also holds true at $z^*_{\gamma \uparrow}$.

For the case of multicast min-cost Program (T1D) for the target flow
$F \in [0,F^*]$, without loss of generality let us assume that
\begin{equation}\label{T1eq29}
 \lambda^{1}_{{k'}}+\lambda^{2}_{{k'}}= \max_{j \in [1,\tau-1]} (\lambda^{1}_{{j}}+\lambda^{2}_{{j}}).
\end{equation}
 From Equation~(\ref{Sec2eq1}) cost of sending the flow $\epsilon >0$ over the path
$l_{k'}$ is given by
\begin{equation}\label{T1eq30}
P^{s}_{T^{{k'}}_1,\epsilon}+P^{r}_{T^{{k'}}_2,\epsilon}
=g^{-1}\left(\frac{\epsilon}{\lambda^{1}_{k'}}\right)+g^{-1}\left(\frac{\epsilon}{\lambda^{2}_{k'}} \right),
 \end{equation}
where $g^{-1}$ is the inverse function of the power function $g$ and
is increasing and concave. From Equation~(\ref{T1eq29}) we get
\begin{equation}\label{T1eq31}
P^{s}_{T^{{k'}}_1,\epsilon}+P^{r}_{T^{{k'}}_2,\epsilon}=
\displaystyle\min_{j \in
[1,\tau-1]}(P^{s}_{T^{{j}}_1,\epsilon}+P^{r}_{T^{{j}}_2,\epsilon}).
 \end{equation}

For a given position of relay $z_r \in \mathcal{C}$, then clearly
$\min(\lambda^{1}_{{k'}},\lambda^{2}_{{k'}})=\min(\lambda^{1}_{{k}},\lambda^{2}_{{k}})$,
i.e. the cheapest path through the relay is that path with the
highest min-cut. This is true because
\begin{equation}\label{T1eq31A}
\begin{split}
\min(\lambda^k_{1},\lambda^k_{2})&=\displaystyle\max_{j \in [1,\tau -1]} (\min(\lambda^j_{1},\lambda^j_{2}))\\
&=\frac{1}{\lambda^k_{1}}+\frac{1}{\lambda^k_{2}}=\displaystyle\min_{j \in [1,\tau -1]} \left(\frac{1}{\lambda^j_{1}}+\frac{1}{\lambda^j_{2}} \right),
\end{split}
\end{equation}
if $\lambda^k_{2}=\max_{j \in [1, \tau-1]} (\lambda^j_{2})$ and this
can be safely assumed for the path with the highest min-cut.

From Equation~(\ref{T1eq31}) and Proposition~\ref{T1P3}, we infer
that $l'_{k}$ is the cheapest path for a unit flow among all the
paths $l_{j} \in [1,\tau-1]$. Moreover from
Equation~(\ref{T1eq31A}), paths $l_{k'} (=l_{k})$ and $l_{\tau}$ can
carry any target multicast flow $F \in [0,F^*]$. So we get four
cases
\begin{enumerate}[(1)]
 \item $P^{s}_{T^{{k'}}_1,\epsilon}+P^{r}_{T^{{k'}}_2,\epsilon} \leq
P^{s}_{T,\epsilon}$ and $\min(\lambda^{1}_{{k'}},\lambda^{2}_{{k'}})
\geq \lambda_{\tau}$,
 \item $P^{s}_{T^{{k'}}_1,\epsilon}+P^{r}_{T^{{k'}}_2,\epsilon} \leq
P^{s}_{T,\epsilon}$ and $\min(\lambda^{1}_{{k'}},\lambda^{2}_{{k'}})
< \lambda_{\tau}$,
\item $P^{s}_{T^{{k'}}_1,\epsilon}+P^{r}_{T^{{k'}}_2,\epsilon} >
P^{s}_{T,\epsilon}$ and $\min(\lambda^{1}_{{k'}},\lambda^{2}_{{k'}})
\geq \lambda_{\tau}$,
\item $P^{s}_{T^{{k'}}_1,\epsilon}+P^{r}_{T^{{k'}}_2,\epsilon} >
P^{s}_{T,\epsilon}$ and $\min(\lambda^{1}_{{k'}},\lambda^{2}_{{k'}}) > \lambda_{\tau}$.
\end{enumerate}

In all of the above cases, the target flow $F \in [0,F^{*}]$ flows
over the paths $l_{k'}$ and $l_{\tau}$ only. Thus, we conclude that
for any relay position $z_r \in \mathcal{C}$ the optimal min-cost
target multicast flow $F$ flows over at most two paths $l_{k'}$ and
$l_{\tau}$, and consequently also at the relay position $z^*_{F
\downarrow}$. This completes the proof.
\end{IEEEproof}

%------------------------------------------------------------------------%
\section{Proof of Proposition~\ref{P2} }\label{ap:Proof-P2}
%------------------------------------------------------------------------%
\begin{IEEEproof}
From Theorem~\ref{T1}, we know that the maximized multicast flow
goes over at most two paths, namely path $\hat{l}_1$ having the
highest min-cut among the paths through $r$ and path $\hat{l}_2$
that spans all the nodes in the system. Moreover, there always exist
at least one relay position such that the min-cut of the path
$\hat{l}_1$ is at least as that of path $\hat{l}_2$, implying that
this also holds at the optimal relay position $z^*_{\gamma
\uparrow}$ solving $(s,T,\mathcal{Z},\gamma)$. This is also true for
point $p^*$ because at $p^*$
\begin{equation*}
\displaystyle\max (h(D_{sp^*}),h(D_{p^*t_n})) \leq h(D_{st_n}),
\end{equation*}
that comes from its definition in Equation~(\ref{Sec2eq10}).

Positioning the relay at $p^*$ will render the highest min-cut of
path $\hat{l}_1$ compared to that for any other position. This is
true from the definition of point $p^*$ itself. If at point $p^*$ we
have $g(\nu)h(D_{sp^*})=D_{p^*}=\min_{i\in
[1,n]}g(\mu)h(D_{p^*t_{i}})$, then
\begin{align*}
F^*_{p^*}&=\min
\left(\frac{g(\mu)}{h(\pi^{p^*}_s)},\frac{g(\nu)}{h(\pi^{p^*}_r)}
\right),
\end{align*}
where, $\pi^{p^*}_s=D_{sp^*}$ and $\pi^{p^*}_r=\max_{i \in
[1,n]}(D_{p^* t_i})$.

From our assumption at point $p^*$ the maximized flow $F^*$ consumes
all the source and the relay powers. Since we only consider those
positions of relay at which the min-cut of path $\hat{l}_1$ is
higher compared to path $\hat{l}_2$, positioning the relay at any
point $p \in \mathcal{C}$ such that $\pi^p_s > \pi^{p^*}_s$ only
renders decreased maximum rate over the hyperarc $C^s_{T_1}$ of the
path $\hat{l}_1$. Implying that $F^*_{p} \leq F^*_{p^*}$, even
though there might be some relay power left.

On the other hand, positioning the relay at point $p$ such that
$\pi^p_s < \pi^{p^*}_s$, increases the maximum rate over the
hyperarc $C^s_{T_1}$, as
\begin{equation*}
h(\pi^p_{s}) \leq h(\pi^{p^*}_{s}) \Longrightarrow
\frac{g(\mu)}{h(D_{sp})} \geq \frac{g(\mu)}{h(D_{sp^*})}.
\end{equation*}

Moreover, we get $\pi^p_r > \pi^{p^*}_r$, as moving away in any
direction from point $p^*$ increases $\max_{j \in T_2}(D_{p^*
t_j})$. Therefore the multicast flow is at this point is given by,
\begin{equation*}
F^*_p= \frac{g(\nu)}{h(\pi^p_r)} + \frac{g\left(\mu -
g^{-1}\left(\frac{g(\nu)h(\pi^p_s)}{h(\pi^p_r)}
\right)\right)}{h(D_{st_n})},
\end{equation*}
where the first term on the right hand side is the flow over the
path $\hat{l_1}=\{C^s_{T_1},C^r_{T_2}\}$ that is limited the
hyperarc $C^r_{T_2}$ and the second term is the flow over the path
$\hat{l_2}=\{C^s_{T}\}$ such that $F^*_p$ is achived by maximizing
the flow over the paths $\hat{l_1}$ and $\hat{l_2}$ successively.

As a Corollary  of Proposition~\ref{T1P1}, it can be seen that
\begin{equation*}
 a_1g(x_1)+a_2g(x_2) \leq a_3g(x_1+x_2),
\end{equation*}
where $g$ is an increasing convex function, $a_i \geq 0$ for $i\in
[1,3]$ are some constants such that $a_3 \geq \max(a_1,a_2)$. This
implies that for any source power $\epsilon >0$, the flow over the
path $\hat{l_1}$ corresponding to the relay position $p^*$ will
always be larger than the sum flow over the paths $\hat{l_1}$ and
$\hat{l_2}$ corresponding to the relay position $p$. Therefore, for
any such relay position $p$, $F^*_{p} \leq F^*_{p^*}$. This proves
the proposition.
\end{IEEEproof}

% \begin{figure}[tp]
% \begin{center}
% \psfrag{a}{{$(a)$}} \psfrag{b}{{$(b)$}}
% \psfrag{s}{{$s$}} \psfrag{t1}{{$t_1$}}
% \psfrag{t2}{{$t_2$}} \psfrag{t3}{{$t_3$}}
% \psfrag{r}{{$r$}} \psfrag{csr}{{$C^s_r$}}
% \psfrag{crs}{{$C^r_s$}} \psfrag{f2}{{$f(x_{2})$}}
% \psfrag{f3}{{$f(x_{3})$}}
% \includegraphics[width=1\columnwidth]{Pictures/ProofP2P3.eps}
% \end{center}
% \vspace{-4mm} \caption{{$r$ can always be placed on the line segment
% $s-t_3$ such that we get in (a): $\min(\pi_s,\pi_r) \leq D_{st_3}
% \Rightarrow \min(\lambda_{T_{1}},\lambda_{T_{2}}) \geq \lambda_{T}$
% (where the path $\{(s,T_1),(r,T_2)\}=\{C^s_r,C^r_{t_3}\}$), and in
% (b): $g^{-1}(h(\pi_s))+g^{-1}(h(\pi_r)) \leq g^{-1}(h(D_{st_3}))$.
% Here, $g^{-1}(h(D)) $ is convex and increasing in $D$, so for any
% point $r$ on $s-t_3$ we get $g^{-1}(h(D_{sr}))+g^{-1}(h(D_{rt_3}))
% \leq g^{-1}(h(D_{st_3}))$. }} \label{fig:ProofP2P3} \vspace{-6mm}
% \end{figure}

%------------------------------------------------------------------------%
\section{Proof of Theorem~\ref{T2} }\label{ap:Proof-T2}
%------------------------------------------------------------------------%
Refer Figure~\ref{fig:Fig2} as a reference example along with the proof. \vspace{1mm}

\begin{IEEEproof}
Let the optimal relay positions $z^{*}_{\gamma \uparrow}$ and
$z^{*}_{F \downarrow}$ be given that solve the problems
$(s,T,\mathcal{Z},\gamma)$ and $(s,T,\mathcal{Z},\gamma,F)$,
respectively. Then the hypearcs of the path
$\hat{l}^\uparrow_{1}=\{C^{s}_{T^\uparrow_1},C^{r}_{T^\uparrow_2}\}$
and $\hat{l}_{1}
^{\downarrow}=\{C^{s}_{T^\downarrow_1},C^{r}_{T^\downarrow_2}\}$ can
be constructed simply by forming the first hyperarcs
$C^{s}_{T^\uparrow_1}$ and $C^{s}_{T^\downarrow_1}$ with radii
$\pi^{\uparrow*}_{s}=D_{s z^{*}_{\gamma \uparrow}}$ and
$\pi^{\downarrow*}_{s}=D_{s z^{*}_{F \downarrow}}$, respectively.
Here, the paths $\hat{l}^\uparrow_{1}$ and $\hat{l}^\downarrow_{1}$
represent the path $\hat{l}_{1}$ corresponding to the relay
positions $z^{*}_{\gamma \uparrow}$ and $z^{*}_{F \downarrow}$,
respectively. Compute the points
\begin{equation*}
\begin{split}
z_{r^\uparrow}= \displaystyle\argmax_{r \in C^{s}_{T^\uparrow_1},
\mathcal{Z} \in C_{\cup}} (\displaystyle\max_{t \in \widehat{T}^{\uparrow}}(D_{rt})), \\
z_{r^\downarrow}= \displaystyle\argmax_{r \in
C^{s}_{T^\downarrow_1}, \mathcal{Z} \in C_{\cup}}
(\displaystyle\max_{t \in \widehat{T}^{\downarrow}}(D_{rt})),
\end{split}
\end{equation*}
where $\widehat{T}^{\uparrow}=\{t \in T| D_{st_j} >
\pi^{\uparrow*}_{s}=D_{sz^{*}_{\gamma \uparrow}}\}$ and
$\widehat{T}^{\downarrow}=\{t \in T| D_{st} >
\pi^{\downarrow*}_{s}=D_{sz^{*}_{F \downarrow}}\}$.

If the points $r^\uparrow$ and $r^\downarrow$ are not the same as
$z^{*}_{\gamma \uparrow}$ and $z^{*}_{F \downarrow}$, respectively,
then this contradicts the optimality of the two points
$z^{*}_{\gamma \uparrow}$ and $z^{*}_{F \downarrow}$. This is true
because the only way to either maximize the rate or minimize the
cost over the hyperarcs $C^{r}_{T^\uparrow_2}$ and
$C^{r}_{T^\downarrow_2}$ is to compute the points inside the
hyperarcs $C^{s}_{T^\uparrow_1}$ and $C^{s}_{T^\downarrow_1}$ that
minimize the maximum among the distances to all the destination
nodes outside these hyperarcs from itself, respectively. Therefore,
the optimal relay positions $z^{*}_{\gamma \uparrow}$ and $z^{*}_{F
\downarrow}$ solving the problems $(s,T,\mathcal{Z},\gamma)$ and
$(s,T,\mathcal{Z},\gamma,F)$, are the points of type $\widehat{r}$
on the curve $\widehat{\mathcal{R}}$. Hence, $z^{*}_{F
\downarrow}=z^{*}_{\gamma' \uparrow}$, for some $\gamma' \in
(0,\infty)$.

Now let us consider that the position $z^*_{1 \downarrow}$ that
minimizes the cost of unit flow (normalized, if the flow values are
less than unity) from $s$ to $T$, and $z^*_{1 \downarrow} =
z^*_{\overline{\gamma} \uparrow}$ for some $\widehat{\gamma} \in
(0,\infty)$. Without loss of generality, let us assume that $z^*_{1
\downarrow}$ is situated on the right of $z^*_{\gamma \uparrow}$ on
the segment $\widehat{\mathcal{R}}$. This implies that the rate over
the source hyperarc $R^{s,z^*_{\widehat{\gamma} \uparrow}}_{T_1}$ is
the limiting term in
\begin{equation*}
F^*_{\hat{l}^{z^*_{\widehat{\gamma} \uparrow}}_1}
=\min(R^{s,z^*_{\widehat{\gamma}
\uparrow}}_{T_1},R^{r,z^*_{\overline{\gamma} \uparrow}}_{T_2}),
\end{equation*}
where $F^*_{\hat{l}^{z^*_{\overline{\gamma} \uparrow}}_1}$ is the
maximized flow over the path $\hat{l}^{z^*_{\overline{\gamma}
\uparrow}}_1$ corresponding to the relay position
$z^*_{\overline{\gamma} \uparrow}$. Furthermore, the only way to
increase the min-cut of the path $\hat{l}_1$ is to position the
relay on the left of $z^*_{\overline{\gamma} \uparrow}$ (closer to
$s$ and $z^*_{\gamma \uparrow}$) on the segment
$\widehat{\mathcal{R}}$, as  positioning the relay further on the
right of $z^*_{\overline{\gamma} \uparrow}$ on the segment
$\widehat{\mathcal{R}}$ will not only increase the cost of unit flow
but will also decrease the min-cut of the path $\hat{l}_1$.
Therefore, for any $F \in [0,F^*]$, $z^*_{F \downarrow} \in
z^*_{\gamma \uparrow} - z^*_{\overline{\gamma} \uparrow}$, where
$z^*_{\gamma \uparrow} - z^*_{\overline{\gamma} \uparrow}$ is the
sub-segment of the piecewise linear segment $\overline{\mathcal{R}}$
joining the point $z^*_{\gamma \uparrow}$ and $z^*_{\widehat{\gamma}
\uparrow}$. The same argument holds for the case when the point
$z^*_{\gamma \uparrow}$ lies on the right of $z^*_{\overline{\gamma}
\uparrow}$. Thus, we can conclude that
\begin{equation*}
  z^{*}_{F \downarrow} = z^{*}_{\widehat{\gamma} \uparrow},
\end{equation*}
for some $\widehat{\gamma} \in
[\min(\overline{\gamma},\gamma),\max(\overline{\gamma},\gamma)]$,
and $\forall F \in [0,F^{*}]$.

Now suppose that for some $\widehat{\gamma}  \in  \gamma
-\overline{\gamma}$ such that $\widehat{\gamma}=\frac{\nu'}{\mu'}$
and $(\mu',\nu') \preceq (\mu,\nu)$, the optimal relay position
maximizing the multicast flow $F$ is given by $z^*_{\widehat{\gamma}
\uparrow} \in  z^*_{\gamma \uparrow} - z^*_{\overline{\gamma}
\uparrow}$ and the maximized multicast flow is given by
$F^*_{z^*_{\widehat{\gamma} \uparrow}}$. Implying that, we need at
least the total source and relay power of value $\mu' + \nu'$ to
achieve the multicast flow $F^*_{z^*_{\widehat{\gamma} \uparrow}}$.
Thus, we can conclude that $z^*_{\widehat{\gamma}
\uparrow}=z^*_{F^*_{z^*_{\widehat{\gamma} \uparrow}} \downarrow}$.
Hence, for any $\widehat{\gamma}  \in  \gamma  -\overline{\gamma}$,
there exist a flow value $F^*_{z^*_{\widehat{\gamma}}} \in [0,F^*]$
such that $z^*_{\widehat{\gamma}
\uparrow}=z^*_{F^*_{z^*_{\widehat{\gamma} \uparrow}} \downarrow}$.
This completes the proof of Theorem~\ref{T2}.
\end{IEEEproof}

\section{Proof of optimality of Max-flow Algorithm }\label{ap:Proof-A1}
%------------------------------------------------------------------------%
We divide the space of max-flow problem in three categories, each
for a step in the Max-flow Algorithm in Figure~\ref{alg:Maxflowalg}.
Proving the optimality for each category will prove the optimality
of the Max-flow Algorithm as any given problem instance
$(s,T,\mathcal{Z},\gamma)$ will fall in one of the three categories.
The first two categories (corresponding to the first two steps of
the Max-flow Algorithm) deal with the instances when at the optimal
relay position $z^*_{\gamma \uparrow}$ all the maximized multicast
flow $F^*$ goes over the single path $\hat{l}_1$. \vspace{1mm}

% \begin{IEEEproof}[Proof of Step $1$]
% For a given problem instance $(s,T,\mathcal{Z},\gamma)$, let us
% maximize the flow $F_{t_n}$ from $s$ to $t_n$ only. Positioning the
% relay at the the point $p$ on the line segment $s-t_n$ such that
% $\frac{g(\mu)}{h(D_{sp})}=\frac{g(\nu)}{h(D_{pt_n})}$ achieves it
% and is denoted by $F^*_{t_n}$. Since $t_n$ is the farthest and the
% limiting node in maximizing the multicast flow from $s$ to $t_n$,
% for any problem instance $(s,T,\mathcal{Z},\gamma)$, the max-flow
% from $s$ to $t_n$ is the upper bound to multicast flow $F$ from $s$
% to $T$.
%
% If all the destination nodes $t_i \in T$ lie in the region
% $C_{\cup}=C^s_{r} \cup C^r_{t_n}$ then all the nodes $t_j \in T$
% will get the maximized flow $F^*_{t_n}$ to $t_n$. Here the two
% circles $C^s_{r}$ and $C^r_{t_n}$ with radii $\pi_s=D_{sp}$ and
% $\pi_r=D_{pt_n}$ are the hyperarcs of the path
% $\hat{l_1}=\{C^s_r,C^r_{t_n}\}$. Hence, $z^*_{\gamma \uparrow}=z_p$
% and $F^*=\frac{g(\mu)}{h(D_{sp})}=\frac{g(\nu)}{h(D_{pt_n})}$.
% \end{IEEEproof}
% \vspace{2mm}

\begin{figure}[tp]
\begin{center}
\psfrag{a}{{$(a)$}} \psfrag{b}{{$(b)$}} \psfrag{c}{{$(c)$}} \psfrag{s}{{\tiny{$s$}}}
\psfrag{t1}{{\tiny{$t_1$}}} \psfrag{t2}{\tiny{$t_2$}} \psfrag{t3}{\tiny{$t_3$}} \psfrag{t4}{\tiny{$t_4$}}
\psfrag{p*}{\tiny{$p^*$}} \psfrag{pt'}{\tiny{$p^*_{T-T'}$}} \psfrag{c1f}{{$C^{t_1}_F$}}
\psfrag{c2f}{{$C^{t_2}_F$}} \psfrag{c3f}{{$C^{t_3}_F$}}
\psfrag{cp}{{$C_p$}} \psfrag{csf}{{$C^s_F$}} \psfrag{p}{{$\hat{p}$}}
\psfrag{cap}{{$C^F_{\cap}$}}
\includegraphics[width=1\columnwidth]{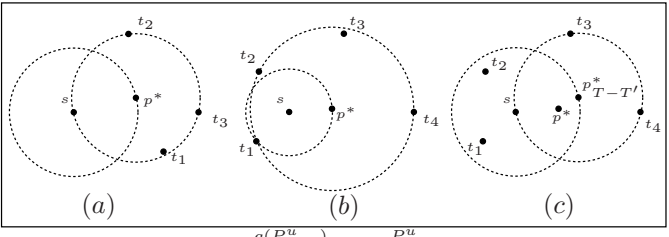}
\end{center}
\vspace{-4mm} \caption{{Consider
$R^{u}_{V_{k_u}}=\frac{g(P^{u}_{V_{k_u}})}{h(D_{uv_{k_u}})}=\frac{P^{u}_{V_{k_u}}}{(D_{uv_{k_u}})^2}$
with $g(\mu)=g(\nu)=1$ in the examples. (a): $|T|=3$ example with
$\frac{1}{(D_{sp^*})^2}=\frac{1}{(D_{p^* t_n})^2}$ and the path
$\hat{l}_1=\{C^s_{T_1},C^r_{T_2}\}$ (corresponding to $p^*$) carries all the maximized flow $F^*$. Therefore, $z^*_{\gamma \uparrow}=z_{p^*}$.
(b): $|T|=4$ example showing the path
$\hat{l}_1=\{C^s_{T_1},C^r_{T_2}\}$ corresponding to $p^*$ but
$\frac{1}{(D_{sp^*})^2}>\frac{1}{(D_{p^* t_4})^2}$. (c): The nodes in $T'=\{t_1,t_2\}$ can be ignored, as the hyperarc $C^s_{T_1}$
of the path $\hat{l}_1=\{C^s_{T_1},C^r_{T_2}\}$ corresponding to
$p^*_{T\backslash T'}$ spans the set $T'$. Thus $z^*_{\gamma \uparrow}=z_{p^*_{T \backslash T'}}$. }}
\label{fig:Maxflowalg1} \vspace{-6mm}
\end{figure}

\begin{IEEEproof}[Proof of Step $1$]
Refer Appendix~\ref{ap:Proof-P2} for the proof of
Proposition~\ref{P2} and Figure~\ref{fig:Maxflowalg1}(a) for an
example.
\end{IEEEproof}

\vspace{2mm}

\begin{IEEEproof}[Proof of Step $2$]
Now assume that $g(\nu)h(D_{sp^*}) < g(\mu)h(D_{p^* t_n})$. This
implies that there exist at least one destination node that is
closer to $s$ than to $p^*$, thus we can build the ordered set
$T'=\{t'_j \in T|D_{st'_j} <
D_{p^{*}t'_j})\}=\{t'_{1},..,t'_{j'}\}$ in increasing
distance from $s$.

We can then recompute the point $p^{*}$ for the destination nodes in
the set $T \backslash {T'}$ and denote it by $p^{*}_{T \backslash
T'}$. If we get, $g(\nu)h(D_{sp^{*}_{T \backslash T'}}) =
g(\mu)h(D_{p^{*}_{T\backslash T'} t_n})$, using Proposition~\ref{P2}
we infer that the point $p^{*}_{T \backslash T'}$ is the optimal
relay position maximizing the mutlicast flow from $s$ to the set of
destinations in the set $T \backslash T'$. Furthermore, if
$D_{st'_{j'}} \leq D_{sp^{*}_{T \backslash T'}}$ then all the nodes
in the set $T \backslash T'$ are spanned by the hyperarc $C^s_{T_1}$
with radii $\pi_s=D_{sp^{*}_{T \backslash T'}}$ corresponding to the
relay position $p^{*}_{T \backslash T'}$. Implying that the
$z^*_{\gamma \uparrow}=z_{p^{*}_{T \backslash T'}}$.
\end{IEEEproof}

Step $2$ essentially gets rid of redundant destinations for
computing the point $p^*$ that are close enough to the source $s$.
Refer Figure~\ref{fig:Maxflowalg1} for an example. \vspace{2mm}

In contrast, if at point $p^{*}_{T \backslash T'}$ we get
$g(\nu)h(D_{sp^{*}_{T \backslash T'}}) <
g(\mu)h(D_{p^{*}_{T\backslash T'} t_n})$, then we can divide the
problem of optimal relay position within two regions in
$\mathcal{C}$. Since the relay position maximizing the multicast
flow in the two regions is going to be unique, we can compare the
two results and declare the global optimal position $z^*_{\gamma
\uparrow}$ solving the problem instance $(s,T,\mathcal{Z},\gamma)$.

First region is the interior of the circle $\mathbf{int}
C^s_{t'_{j'}}$ centered at $s$ with radius $D_{st'_{j'}}$ and the
second region as the area $\mathcal{C} \backslash \mathbf{int}
C^s_{t'_{j'}}$ which is the rest of region in $\mathcal{C}$ minus
the first region. Let us denote the optimal relay positions inside
the first and second regions with $z^*_{1}$ and $z^*_{2}$,
respectively. At first, we present the following two propositions
that will come in handy for the proof of optimality of Step $3$.
Refer Figure~\ref{fig:Maxflowalg2} with $g(P)=P$ and $h(D)=D^2$ for
an example.

\hspace{1mm}
\begin{proposition}\label{P8}
For any $|T|=2$ case max-flow problem instance such that
$g(\nu)h(D_{sp^*}) < \max_{i \in [1,2]}(g(\mu)h(D_{p^*
t_i}))=D_{p^*}$, $z^{*}_{1} \in \mathbf{int} C^s_{t_1}$ lies on the
line segment $p^*$ and $p_{12}$.
\end{proposition}
\begin{IEEEproof}
 Consider the two destination node case
 such that $g(\nu)h(D_{sp^*}) < \max_{i \in [1,2]}(g(\mu)h(D_{p^*
t_i}))$, where
\begin{equation*}
p^*=\displaystyle\argmin_{p \in \mathcal{C}} (\max(g(\nu)h(D_{sp}),g(\mu)h(D_{pt_1}),g(\mu)h(D_{pt_2}))).
\end{equation*}
This implies that $D_{st'_{j'}} > D_{sp^{*}_{T \backslash T'}}$, where
\begin{equation*}
 p^*_{T \backslash T'}= \displaystyle\argmin_{p \in \mathcal{C}} (\max(g(\nu)h(D_{sp}),g(\mu)h(D_{pt_2}))),
\end{equation*}
and  $T'=\{t_1\}$. Moreover, as a consequence of Theorem~\ref{T1},
only those points need to be considered such that positioning the
relay at this point gives higher min-cut for the path $\hat{l_1}$
than that of path $\hat{l_2}$, inside the region $\mathcal{C} \cap
\mathbf{int} C^s_{t_1}$.

The circle $C^s_{t'_{j'}}$ is the circle $C^s_{t_1}$ centered at $s$
with radius $D_{st_1}$. Let the perpendicular bisector of the nodes
$t_1$ and $t_2$ be denoted as $\perp_{12}$, and its intersection
point with the segment $s-t_2$ by $p_{12}$. Now consider any point
$p \in \mathcal{C} \cap \mathbf{int} C^s_{t_1}$ in the halfplane
containing the node $t_2$ of the bisector $\perp_{12}$ and let $p'$
denote the closest point on $\perp_{12}$ to $p$. Then clearly,
$D_{sp'} < D_{sp} \Rightarrow h(D_{sp'}) \leq h(D_{sp})$ and
$D_{p't_1}=D_{p't_2} < D_{pt_1} \Rightarrow h(D_{p't_2}) \leq
h(D_{pt_1})$. This implies that the min-cut of the path $\hat{l_1}$
reduces when the relay is positioned at the point $p$ compared to
that at point $p'$. Therefore, for any relay position in the
halfplane (of $\perp_{12}$) containing $t_2$, there always exist a
position on the segment $p^* -p_{12}$ that is a better candidate for
the optimal relay position maximizing the multicast flow.

\begin{figure}[tp]
\begin{center}
\psfrag{a}{{$(a)$}} \psfrag{b}{{$(b)$}} \psfrag{c}{{$(c)$}} \psfrag{s}{{{$s$}}}
\psfrag{t1}{{{$t_1$}}} \psfrag{t2}{{$t_2$}} \psfrag{t12}{{$\perp_{12}$}} \psfrag{t4}{\tiny{$t_4$}}
\psfrag{p*}{{$p^*$}} \psfrag{p}{{$p$}} \psfrag{p'}{{$p'$}} \psfrag{C}{{$C^{s}_{t_{1}}$}}
\psfrag{t13}{{$\perp_{13}$}} \psfrag{t23}{{$\perp_{23}$}} \psfrag{p12}{{$p_{12}$}}
\psfrag{l}{{$\ell_{st_n}$}} \psfrag{csf}{{$C^s_F$}}
\psfrag{cap}{{$C^F_{\cap}$}}
\includegraphics[width=1\columnwidth]{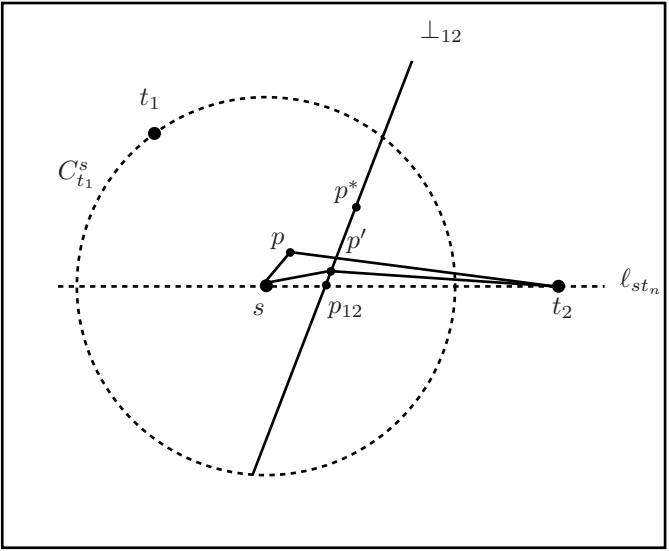}
\end{center}
\vspace{-4mm} \caption{{$|T|=2$ node example with $T'=\{t_1\}$. Note
that, $D_{sp} < D_{sp'}$ and $D_{pt_2} < D_{p' t_2}$. Therefore, the
rate over the paths $\hat{l}^p_1=\{(s,p),(p,t_{1}t_2)\}$ and
$\hat{l}^{p'}_1=\{(s,p'),(p',t_{1}t_2)\}$ is limited by the
hyperarcs $(p,t_{1}t_2)$ and $(p',t_1 t_2)$, respectively.}}
\label{fig:Maxflowalg2} \vspace{-6mm}
\end{figure}

Now consider a point $p \in \mathcal{C} \cap \mathbf{int} C^s_{t_1}$
in the halfplane (of $\perp_{12}$) containing the node $t_1$ such
that $D_{sp} < D_{sp'}$, then we get $D_{pt_2} > D_{p't_2}$. Note
that, positions $p$ for which $D_{sp} > D_{sp'}$ are uninteresting
due to the fact that not only the min-cut of the path $\hat{l}_1$
decreases but also that the maximum possible rate over the hyperarc
$C^s_{T_1}$ decreases compared to the relay position $p'$
(corresponding to $p$). Then the maximized multicast flow for the
relay position $p$ and $p'$ such that $D_{sp} < D_{sp'}$ is given by
\begin{align*}
F^*_{p}&=\frac{g(\nu)}{h(D_{pt_2})}+ \frac{g\left(\mu-g^{-1}\left(\frac{g(\nu)h(D_{sp})}{h(D_{pt_2})}\right)\right)}{h(D_{st_2})}, \\
F^*_{p'}&=\frac{g(\nu)}{h(D_{p't_2})}+
\frac{g\left(\mu-g^{-1}\left(\frac{g(\nu)h(D_{sp'})}{h(D_{p't_2})}\right)\right)}{h(D_{st_2})},
\end{align*}
respectively. The first term in the above equations is the flow
$F_{\hat{l}_1}=\min(R^s_{{T_1}},R^r_{{T_2}})=R^r_{{T_2}}$  over the
path $\hat{l}_1$ which is limited by the rate $R^r_{{T_2}}$ over the
hyperarc $C^r_{T_2}$ and the second term is the flow over the path
$\hat{l}_2$ which is the function of total source power $\mu$ minus
the power used over the hyperarc $C^{s}_{T_1}$. As $D_{p't_2} <
D_{pt_2}$, we have $\frac{g(\nu)}{h(D_{pt_2})} \leq
\frac{g(\nu)}{h(D_{p't_2})}$. Furthermore, from triangle inequality
we get $D_{sp'}-D_{sp} < D_{p't_2}-D_{pt_2}$, implying that
\begin{equation*}
\frac{g(\epsilon)}{h(D_{sp})}-\frac{g(\epsilon)}{h(D_{sp'})} \leq \frac{g(\epsilon)}{h(D_{p't_2})}-\frac{g(\epsilon)}{h(D_{pt_2})},
\end{equation*}
for any $P^s_{T_1}=P^r_{T_2}=\epsilon \geq0$. Therefore, we get more
flow over the path $\hat{l}_1$ corresponding to the relay position
$p'$ compared to that of relay position $p$. Lastly, since the
min-cut of the path $\hat{l}_2$ is strictly smaller than that of
path $\hat{l}_1$, it can be concluded that $F^*_{p} \leq F^*_{p'}$.
Hence, for any point $p \in \mathcal{C} \cap \mathbf{int} C^s_{t_1}$
there exist a point $p'$ on the segment $p^* - p_{12}$ of the
perpendicular bisector $\perp_{12}$, such that $F^*_{p} \leq
F^*_{p'}$, hence $z^*_{1} \in p^*-p_{12}$.
\end{IEEEproof}

\vspace{1mm}
 Note that if $h$ is an strictly increasing function of
distance, then the following inequality would hold strictly $F^*_{p}
< F^*_{p'}$. For $|T| >2$, Proposition~\ref{P8} can be generalized
in the following way. For simplicity, let us assume that the line
$\ell_{st_n}$ passing through $s$ and $t_n$ is horizontal and point
$p^*$ lies above $\ell_{st_n}$ (ref. Figure~\ref{fig:Maxflowalg2}
for example). Now, let the set of perpendicular bisectors for each
pair of nodes in $T$ be denoted by
$\perp=\{\perp_{12},\perp_{13},..,\perp_{n-1n}\}$, where
$\perp_{ab}$ denotes the perpendicular bisector of the nodes $t_a
\in T$ and $t_b \in T$, and $|\perp|=\frac{n!}{n!(n-2)!}$. Most of
the bisectors $\perp_{ab} \in \perp$ will intersect with the line
$\ell_{st_n}$ and let $\angle_{ab}$ denote the angle between the
point $s$, the point of intersection of $\perp_{ab}$ and
$\ell_{st_n}$, and any point on the bisector $\perp_{ab}$ above the
line $\ell_{st_n}$. The closest point on the perpendicular bisector
$\perp_{ab}$ to $s$ is denoted by $p^s_{ab}$. In addition, let
$\perp \supset \overline{\perp}=\{\perp_{\bar{1} n},\perp_{\bar{2}
n},..,\perp_{\bar{m} n}\}$ be the subset of bisectors
$\perp_{\bar{j}n}$ of the nodes $\bar{t}_j \in T$ and $t_n$ (the
farthest node from $s$ in the system) for $j \in [1,m]$, such that
there exist a segment $p_{\bar{j}n}-q_{\bar{j}n}$ of
$\perp_{\bar{j}n}$ in $\mathcal{C}$ so that the farthest nodes from
any point $p \in p_{\bar{j}n}-q_{\bar{j}n}$ are the nodes
$\bar{t}_j$ and $t_n$ themselves.  For example, in
Figure~\ref{fig:Maxflowalg2} the segment $p^* - p_{12}$ of bisector
$\perp_{12}$. Finally, without loss of generality we assume that
$\perp_{\bar{l}n} \in \bar{\perp}$ be the perpendicular bisector
such that the point $p^s_{\bar{l}n}$ lies on the segment
$p_{\bar{l}n}-q_{\bar{l}n}$.

\vspace{2mm}
\begin{proposition}\label{P9}
For $|T|=n >2$ max-flow problem instance such that
$g(\nu)h(D_{sp^*}) < D_{p^*}= \max_{i \in [1,n]}(g(\mu)h(D_{p^*
t_i}))$, $z^{*}_{1} \in \mathcal{C} \cap \mathbf{int}
C^s_{{t'}_{j'}}$ lies on the piecewise linear segment $(p^*
- q_{\bar{1}n},q_{\bar{1}n} - q_{\bar{2}n},.., q_{\bar{l}-1n} -
q_{\bar{l}n})$.
\end{proposition}
\begin{IEEEproof}
Consider the region $\mathcal{C} \cap \mathbf{int} C^s_{{t'}_{j'}}$
for determining the best possible relay position $z^*_1$ maximizing
the multicast flow $F$. $p^*$ is already a good reference point in
$\mathcal{C} \cap \mathbf{int} C^s_{{t'}_{j'}}$. From
Proposition~\ref{P8}, we know that position $p$ such that $D_{sp} <
D_{sp^*}$, i.e. the rate over the hyperarc $C^s_{T_1}$ increases for
a given power $P^s_{T_1}$, is an interesting position in terms of
finding the point $z^*_1$, thus we will only consider such
directions from $p^*$. In other words, only those bisectors
$\perp_{\bar{j}n} \in \overline{\perp}$ need to be considered that
intersect $\ell_{st_n}$ on the segment $s-t_n$ and make an obtuse
angle $\angle_{{\bar{j}n}}$.

Let $t_{\bar{1}}$ be the limiting node in determining $p^*$, i.e.
$g(\mu)h(D_{p^*t_{\bar{1}}})=D_{p^*}$ whose bisector makes the
largest obtuse angle $\angle_{{\bar{1}}n}$ with $\ell_{st_n}$ (e.g.
node $t_1$ in Figure~\ref{fig:Maxflowalg3}(a)). Also, $t_{\bar{1}}$
and $t_n$ are the farthest limiting nodes  from $s$ in determining
the point $p^*$. This implies, that there exist a segment $p^* -
q_{\bar{1}n}$ on the bisector $\perp_{\bar{1}n}$ (towards $s$) such
that $t_{\bar{1}}$ and $t_n$ are the farthest nodes from any point
$p \in p^* - q_{\bar{1}n}$. If $p^s_{\bar{1}n} \in
p^*-q_{\bar{1}n}$, then using Proposition~\ref{P8} for any position
$p \in \mathcal{C} \cap C^s_{t'_{j'}}$ and the closest point
$p'_{\bar{1}n}$ to $p$ on $\perp_{\bar{1}n}$, we get three cases
\begin{align*}
\text{either} \hspace{2mm} D_{sp} < D_{s\bar{p}'_{\bar{1}n}} \hspace{2mm} & \text{and} \hspace{2mm}D_{pt_n} > D_{\bar{p}'_{\bar{1}n}t_{n}}, \\
\text{or} \hspace{2mm} D_{sp} > D_{s\bar{p}'_{\bar{1}n}} \hspace{2mm} & \text{and} \hspace{2mm} D_{pt_{n}} > D_{\bar{p}'_{\bar{1}n} t_{n}}, \\
\text{or} \hspace{2mm} D_{sp} > D_{s\bar{p}'_{\bar{1}n}}
\hspace{2mm} & \text{and} \hspace{2mm} D_{pt_{n}} <
D_{\bar{p}'_{\bar{1}n} t_{2}}.
\end{align*}
Then if $p^s_{\bar{1}n} \in p_{\bar{1}n}-q_{\bar{1}n}$, using
Proposition~\ref{P8} we can deduce that $z^*_{\gamma \uparrow} \in
p^* - q_{\bar{1}n}$. All other points $p\in \mathcal{C} \cap
\mathbf{int} C^s_{{t'}_{j'}}$ such that $p'_{\bar{1}n}$ lies outside
the segment $p^* - q_{\bar{1}n}$ can be ignored.

On the other hand, suppose that $p^s_{\bar{1}n} \notin
p_{\bar{1}n}-q_{\bar{1}n}$. Then, there exist another bisector
$\perp_{\bar{2}n}$ (of the nodes $t_{\bar{2}}$ and $t_n$) that intersects
$\perp_{\bar{1}n}$, say at point $q_{\bar{1}n}$ and contains a segment $q_{\bar{1}n}-q_{\bar{2}n} \in \mathcal{C} \cap \mathbf{int} C^s_{{t'}_{j'}}$
such that for any point $p \in q_{\bar{1}n}-q_{\bar{2}n}$, $t_{\bar{2}}$ and $t_n$ are the farthest nodes in the system. Using proposition~\ref{P8} again, we can infer that
for all the points $p \in \mathcal{C} \cap \mathbf{int} C^s_{{t'}_{j'}}$ such that $D_{pp'_{\bar{1}n}} \leq D_{pp'_{\bar{2}n}}$, positioning the relay at $p'_{\bar{1}n}$
will render $F^*_{p} \leq F^*_{p'_{\bar{1}n}}$. Similarly, if $D_{pp'_{\bar{1}n}} > D_{pp'_{\bar{2}n}}$, then positioning the relay at $p'_{\bar{2}n}$
will render $F^*_{p} \leq F^*_{p'_{\bar{2}n}}$. Finally, if $p^s_{\bar{2}n} \in q_{\bar{1}n}-q_{\bar{2}n}$, then we can conclude that
$z^*_{\gamma \uparrow} \in (p_{\bar{1}n} - q_{\bar{1}n},q_{\bar{1}n} - q_{\bar{2}n})$.

\begin{figure}[tp]
\begin{center}
\psfrag{a}{{$(a)$}} \psfrag{b}{{$(b)$}} \psfrag{c}{{$(c)$}}
\psfrag{s}{{{$s$}}} \psfrag{t1}{{{$t_1$}}} \psfrag{t4}{{{$t_4$}}}
\psfrag{t5}{{{$t_5$}}} \psfrag{t2}{{$t_2$}} \psfrag{t3}{{$t_3$}}
\psfrag{t4}{{$t_4$}} \psfrag{p*}{\tiny{$p^*$}} \psfrag{p13}{\tiny{$p_{\bar{1}3}$}}
\psfrag{12}{{$\perp_{12}$}} \psfrag{23}{{$\perp_{23}$}} \psfrag{ab}{\tiny{$\ell^{ \bar{12}}_{\angle}$}}
\psfrag{13}{{$\perp_{13}$}}
\psfrag{15}{{$\perp_{15}$}}\psfrag{s13}{\tiny{$p^s_{\bar{1}3}$}} \psfrag{lst}{{$\ell_{st_3}$}}
\includegraphics[width=1\columnwidth]{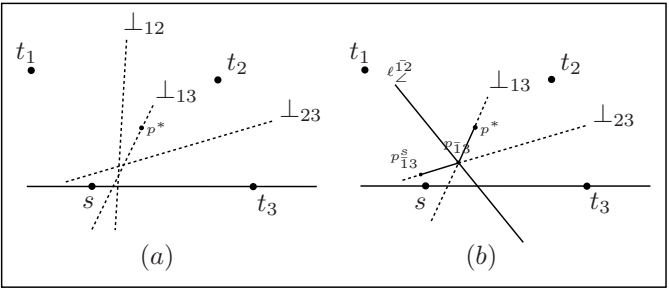}
\end{center}
\vspace{-4mm} \caption{{ $|T|=3$ example with $T'=\{t_1\}$. (a):
$\perp=\{\perp_{12},\perp_{13},\perp_{23}\}$. (b):
$\bar{\perp}=\{\perp_{\bar{1}3},\perp_{\bar{2}3}\}$, where
$t_{{\bar{1}}}=t_1$ and $t_{{\bar{2}}}=t_2$. $\ell^{
\bar{12}}_{\angle}$ is the angle bisector of $\angle_{p^*
p_{\bar{1}3} p^s_{\bar{2}3}}$ dividing the region $\mathcal{C} \cap
C^s_{t_1}$ into the two halfplanes of points that are closer to
segments $p_{\bar{1}3}-q_{\bar{1}3}$ and
$q_{\bar{1}3}-q_{\bar{2}3}$. }} \label{fig:Maxflowalg3}
\vspace{-6mm}
\end{figure}

Generalizing to case of $l$ such bisectors such that $p^s_{\bar{l}n}
\in q_{\bar{l}-1n} - q_{\bar{l}n}$, we conclude that $z^*_{1} \in
(p_{\bar{1}n} - q_{\bar{1}n},
q_{\bar{1}n}-q_{\bar{2}n},..,q_{\bar{l}-1n}-q_{\bar{l}n})$. The same
argument would suffice if $p^*$ lies below the line $\ell_{st_n}$
and this completes the proof.
\end{IEEEproof}

\vspace{1mm} Now we are ready to prove the optimality of Step $3$ of
the Max-flow Algorithm.

\begin{IEEEproof}[Proof of Step $3$] Following Step $2$, assume that not all
the nodes in the set $T'$ lie inside the hyperarc $C^s_{T_1}$ with
radius $\pi_s=D_{sp^{*}_{T-T'}}$, then by reforming the hyperarc
$C^s_{T_1}$ with radius $\pi_s=D_{st'_{j'}}$ we can compute the
point
\begin{equation}
\hat{p}=\displaystyle\argmin_{p \in C^s_{T_1}} (\displaystyle\max_{t_j \in T \backslash \{T'\}}(D_{p t_j})).
\end{equation}
The point $\hat{p}$ will always lie on the circumference of the
hyperarc circle $C^s_{T_1}$. Positioning the relay $r$ at $\hat{p}$
gives the hyperarc $C^r_{T_2}$ with radius $\pi_r=D_{\hat{p}t_n}$,
thus rendering the path $\hat{l}_1=\{C^s_{T_1},C^r_{T_2}\}$ for the
relay position $\hat{p}$. Since $D_{sp^{*}_{T-T'}} <
D_{st'_{j'}}=D_{s\hat{p}}$, this implies $g(\nu)h(D_{s\hat{p}}) \geq
g(\mu)h(D_{\hat{p} t_n})$. Thus maximizing the rate over the path
$\hat{l}_{1}$ will alone maximize the multicast flow $F$ for the
position $\hat{p}$, giving
\begin{equation}
F^{*}_{\hat{p}}=\min \left(\frac{g(\mu)}{h(D_{s\hat{p}})},\frac{g(\nu)}{h(D_{\hat{p}t_n})} \right).
\end{equation}

For any relay position $p$ with $z_{p} \in \mathcal{C} \backslash \textbf{int} C^s_{T_1}$  the
min-cut of the path $\hat{l}_1$ reduces compared to the position
$\hat{p}$ simply because $D_{s\hat{p}} < D_{sp}$, implying
that the maximized multicast rate $F^*_{p} \leq F^*_{\hat{p}}$.
Therefore, $z^*_2=z^*_{\hat{p}}$.

For all the positions inside the circle $C^s_{t'_{j'}}$, from
Proposition~\ref{P9}, we know that $z^*_{1} \in (p^* -
q_{\bar{1}n}, q_{\bar{1}n} - q_{\bar{2}n},.., q_{\bar{l}-1n} -
q_{\bar{l}n})$, where $(p^* -
q_{\bar{1}n}, q_{\bar{1}n} - q_{\bar{2}n},.., q_{\bar{l}-1n} -
q_{\bar{l}n})$ is piecewise linear segment made up of the sub-segments of perpendicular
bisectors $\overline{\perp}=\{\perp_{\bar{1}n},..,\perp_{\bar{l}n}\}$,
where from any point on the bisector $\perp_{\bar{k}n}$,
$t_{\bar{k}}$ and $t_{n}$ are the farthest nodes in the system, for
all $k \in [1,l]$. Let us re-denote this piecewise linear segment by $s=(s_{\bar{1}n},..,s_{\bar{l}n})$, where
$s_{\bar{k}n}$ denotes the sub-segment $q_{\bar{k}-1n} -
q_{\bar{k}n}$ for $k \in [1,n]$.

It is easy to compute the equation of each bisector in
$\overline{\perp}$. Let the equations be given by
\begin{equation}\label{mfa10}
y=m_{\bar{1}}x+ c_{\bar{1}}, \hspace{2mm} \forall k \in [1,l],
\end{equation}
where $y=m_{\bar{1}}x+ c_{\bar{1}}$ is the equation of the bisector
$\perp_{\bar{k}n}$ and the segment $q_{\bar{k}-1n} - q_{\bar{k}n}$
can be represented by limiting the values of $x \in
[x_{q_{\bar{k}-1}},x_{q_{\bar{k}}}]$, where
$z_{q_{\bar{k}-1n}}=(x_{q_{\bar{k}-1n}},y_{q_{\bar{k}-1n}})$ are the
coordinates of the point ${q_{\bar{k}-1n}}$ in the plane. Moreover,
as we know that for any point $p_{\bar{k}} \in
{q_{\bar{k}-1n}}-{q_{\bar{k}n}}$, the limiting hyperarc is
$C^r_{T_2}$, implying that the maximized multicast flow
$F^*_{p_{\bar{k}}}$ can be achieved simply by maximizing the flows
over the paths $\hat{l}_1$ and $\hat{l_2}$ in succession, we get
\begin{equation*}
F^*_{p_{\bar{k}}}=\frac{g(\nu)}{h(D_{p_{\bar{k}}t_{\bar{k}}})}+\frac{
g\left(\mu - g^{-1}\left(
\frac{h(D_{sp_{\bar{k}}})g(\nu)}{h(D_{p_{\bar{k}}t_{\bar{k}}})}
\right)\right)}{h(D_{st_n})},
\end{equation*}
where the only variables are $D_{sp_{\bar{k}}}$ and
$D_{p_{\bar{k}}t_{\bar{k}}}$ as $p_{\bar{k}} \in {q_{\bar{k}-1n}}-{q_{\bar{k}n}}$. The variables $D_{sp_{\bar{k}}}$ and
$D_{p_{\bar{k}}t_{\bar{k}}}$ further  are  functions of coordinates
$(x_{p_{\bar{k}}},y_{p_{\bar{k}}})$ of point $p_{\bar{k}}$. Using
the Equation (\ref{mfa10}), we can rewrite
$F^*_{p_{\bar{k}}}=F^*_{\bar{k}}(x_{\bar{k}})$ as a function of
single variable $x_{\bar{k}}$, where $x_{{\bar{k}}} \in
[x_{q_{\bar{k}-1}},x_{q_{\bar{k}}}]$. Then the optimal relay
position maximizing the multicast flow $F$ in the region
$\mathcal{C} \cap \mathbf{int} C^s_{t'_{j'}}$ is given by $z_1^*=(x^*_{1},y^*_{1})$, where
\begin{equation}\label{mfa11}
x^*_{1}=\displaystyle\argmax_{x_{\bar{k}}}
(\displaystyle\max_{\bar{k} \in [\bar{1},\bar{l}]}
F^*_{\bar{k}}(x_{{\bar{k}}})),
\end{equation}
where $x_{{\bar{k}}} \in [x_{q_{\bar{k}-1}},x_{q_{\bar{k}}}], \forall
\bar{k} \in [\bar{1},\bar{l}]$ and $y^*_{1}=m_{\bar{k}}x^*_{\bar{k}}+ c_{\bar{k}}$.

Finally, comparing the values of $F^*_{1}$ and $F^*_{2}$,
the optimal relay position solving $(s,T,\mathcal{Z},\gamma)$ is
given by
\begin{equation*}
z^*_{\gamma \uparrow}=\begin{cases} z^*_{1} & \text{if $F^*_{1}
> F^*_{2}$.}\\
z^*_{2} & \text{if $F^*_{1} < F^*_{2}$.}
\end{cases}
\end{equation*}
This completes the proof of optimality of Step $3$ of Max-flow
Algorithm.
\end{IEEEproof}
\vspace{1mm} \emph{Remark on solving Equation~ (\ref{mfa11}):} The
single variable function $F^*_{\bar{k}}(x_{{\bar{k}}})$ is
non-convex and smooth over the domain $x_{{\bar{k}}} \in
[x_{q_{\bar{k-1}}},x_{q_{\bar{k}}}], \forall \bar{k} \in
[\bar{1},\bar{l}]$. Moreover it can be proven that there exist a
single stationary point (maxima) of the function in the domain $[x_{q_{\bar{k}-1}},x_{q_{\bar{k}}}]$.
Therefore, using gradient based approach the global maxima can be
achieved, implying that non-convexity is not a hinderance.

%------------------------------------------------------------------------%
\section{Proof of optimality of Min-cost Algorithm }\label{ap:Proof-A2}
%------------------------------------------------------------------------%
Assume that for a given problem instance
$(s,T,\mathcal{Z},\gamma,F)$ the Inequality~(\ref{Sec3Beq4}) holds and at the optimal relay position $z^*_{F \downarrow}$ all the target flow $F$ goes over
path $\hat{l}_1$ only.
Refer Figures~\ref{fig:Mincostalg1}-\ref{fig:Mincostalg2} for
example.

\hspace{1mm}

\begin{IEEEproof}[Proof of Step $2$]
Assume that at the optimal relay
position, all the min-cost multicast flow will go over path
$\hat{l}_2$ only. This implies that the intersection region
$C'_{\cap}=C'^s \cap C'^{t_n}$ of the circles $C'^s$ and $C'^r$ with
radii $\pi'_s=h^{-1}\left( \frac{g(\mu)}{F} \right)$ and
$\pi'_{t_n}=h^{-1}\left( \frac{g(\nu)}{F} \right)$, respectively,
contains at least one point, and this assures feasibility.
Positioning the relay at any point inside $C'_{\cap}$ will fetch the
multicast flow of value $F$ over the path $\hat{l}_2$. Implying that
the relay position in $C'_{\cap}$ minimizing the cost of unit flow
also minimizes the cost of flow $F$ and is thus the optimal relay
position solving $(s,T,\mathcal{Z},\gamma,F)$.

Now consider the point
\begin{equation}\label{A2eq1}
\widehat{p} =
\displaystyle\argmin_{p \in C'_{\cap}} (h(D_{sp}) +
\displaystyle\max_{i \in [1,n]}(h(D_{pt_i}))).
\end{equation}
It can be seen that all the nodes in the set
$\widehat{T}=\{\widehat{t} \in T| D_{s{\widehat{t}}} \leq
D_{s{\widehat{p}}}\}$ can be ignored as they are close enough to the
source node to be spanned by the source hyperarc $C^s_{T_1}$ of the path
$\hat{l}_1$. Therefore, we can recompute
\begin{equation*}
\widehat{p} = \displaystyle\argmin_{p \in C'_{\cap}} (h(D_{sp}) +
\displaystyle\max_{t \in T \backslash \{\hat{T}\}}(h(D_{pt_i}))),
\end{equation*}
and get rid of unnecessary bias. Denote the cost of unit flow
$\Psi_{\widehat{p}}=h(D_{s\widehat{p}})+h(D_{\widehat{p} t_n})$
corresponding to the relay position $\widehat{p}$.

Now consider the region $C'_{\cap} \cap C'^s_{\widehat{p}}$, where  the circle $C'^s_{\widehat{p}}$ has the
radius $D_{s{\widehat{p}}}$. $\widehat{p}$ is
the optimal relay position in the region minimizing the cost of flow $F$, as all the destination nodes inside the
circle $C'^s_{\widehat{p}}$ are not the limiting nodes determining
the position $\widehat{p}$. But the point $\widehat{p}$ is not
necessarily the global optimum for the whole region $C'_{\cap}$.
So we break the problem of finding the global optimal relay position
minimizing the cost of multicast flow $F$, into finding the optimal
relay position among disjoint regions of $C'_{\cap}$ and then
compare them to declare the global optimal relay position.

\begin{figure}[tp]
\begin{center}
\psfrag{cs}{{$C'^s$}} \psfrag{cn}{{$C'^{t_4}$}}
\psfrag{p}{{$\widehat{p}$}} \psfrag{s}{{{$s$}}}
\psfrag{t1}{{{$t_1$}}} \psfrag{t4}{{{$t_4$}}} \psfrag{t5}{{{$t_5$}}}
\psfrag{t2}{{$t_2$}} \psfrag{t3}{{$t_3$}} \psfrag{t4}{{$t_4$}}
\psfrag{p*}{\tiny{$p^*$}} \psfrag{p13}{\tiny{$p_{\bar{1}3}$}}
\psfrag{12}{{$\perp_{12}$}} \psfrag{23}{{$\perp_{23}$}}
\psfrag{ab}{\tiny{$\ell^{ \bar{12}}_{\angle}$}}
\psfrag{13}{{$\perp_{13}$}}
\psfrag{15}{{$\perp_{15}$}}\psfrag{s13}{\tiny{$p^s_{\bar{1}3}$}}
\psfrag{lst}{{$\ell_{st_3}$}}
\includegraphics[width=0.6\columnwidth]{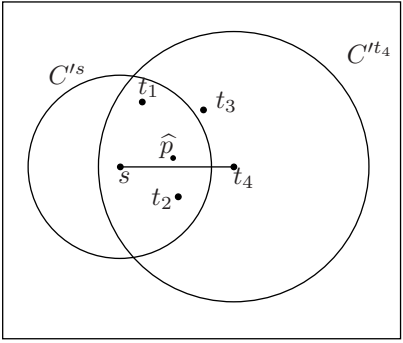}
\end{center}
\vspace{-4mm} \caption{{ Consider the $|T|=4$ node system, with
$C'_{\cap}=C'^s \cap C'^{t_4}$ and the point $\widehat{p}$ that is
the optimal relay position minimizing the cost of unit flow over the
path $\hat{l}_1$ in the region $C'_{\cap} \cap
C'^s_{\widehat{p}}$.}} \label{fig:Mincostalg1} \vspace{-6mm}
\end{figure}

Denote the set of nodes $\overline{T}=\{t \in T \backslash
\{\widehat{T},t_n\}|D_{st}>\pi^{\widehat{p}}_s, D_{st} \leq
\pi'_s\}=\{\overline{t}_1,.., \overline{t}_l\}$, where
$\pi^{\widehat{p}}_s=D_{s{\widehat{p}}}$. Consider finding the
optimal relay position $\overline{p}_1$ in the region $C'_{\cap}
\cap (\overline{C}^s_{2} \backslash \overline{C}^s_{1})$ that
minimizes the cost of unit flow over the path $\hat{l}_1$, where
circles $\overline{C}^s_{1}$ and $\overline{C}^s_{2}$ are centered at
$s$ with radii $D_{s\overline{t}_1}$ and $D_{s\overline{t}_2}$,
respectively. Then the problem can be stated as, {\footnotesize{
\begin{equation}\label{A2eq2}
\widehat{p}_1 = \displaystyle\argmin_{p \in \overline{C}^s_{1}}
(\max(h(D_{sp}),h(D_{s\overline{t}_1})) + \displaystyle\max_{t \in
\overline{T}_1}(h(D_{pt}))),
\end{equation}}}where the set
$\overline{T}_1=\{t \in T|D_{st} > D_{s\overline{t}_1}\}$ consists
of destination nodes lying outside the circle $\overline{C}^s_{1}$.
The Program in Equation~(\ref{A2eq2}) outputs for the optimal relay
position minimizing the cost of unit flow over the hyperarc
$\hat{l}_1$ for the relay position in the region $C'_{\cap} \cap
(\overline{C}^s_{2} \backslash \overline{C}^s_{1})$. Alhtough, the
region $C'_{\cap} \cap (\overline{C}^s_{2} \backslash
\overline{C}^s_{1})$ is non-convex, the optimization program in
Equation~(\ref{A2eq2}) is convex and easy to solve. Similarly, for
all the nodes in the set $\overline{t}_j \in \overline{T}$, we can
compute {\footnotesize{
\begin{equation}\label{A2eq3}
\widehat{p}_j = \displaystyle\argmin_{p \in \overline{C}^s_{j}}
(\max(h(D_{sp}),h(D_{s\overline{t}_{j-1}})) + \displaystyle\max_{t
\in \overline{T}_j}(h(D_{pt}))),
\end{equation}}}and the cost of unit flow
at the optimal relay positions $\widehat{p}_j$ in the disjoint
region $C'_{\cap} \cap (\overline{C}^s_{j+1} \backslash
\overline{C}^s_{j})$ can be calculated, thus denote the vector
$\overline{\Psi}=\{\overline{\Psi}_1,..,\overline{\Psi}_l\}$.
Furthermore, generating the set $\overline{\Psi}$ needs $<|T|=n$
iterations. Finally, we get

\begin{equation*}
z^*_{F \downarrow}=
\begin{cases}
z_{\widehat{p}} & \text{if $\Psi_{\widehat{p}} \leq \overline{\Psi}_m$,}\\
z_{\overline{p}_m} & \text{if $\Psi_{\widehat{p}} \geq
\overline{\Psi}_m$},
\end{cases}
\end{equation*}
where, $\overline{\Psi}_m = \min_{j \in [1,l]}(\overline{\Psi}_k)$.

Note that, the dividing of the region $C'_{\cap}$ into the disjoint
non-convex regions $C'_{\cap} \backslash \mathbf{int}
\overline{C}^s_{j}$ lets capture the idea of positioning the relay
anywhere in the region  $C'_{\cap}$ such that the source hyperarc
$C^s_{T_1}$ of the path $\hat{l}_1$ spans all the destination nodes
that are closer to source than relay and this ensures the global
optimality. This completes the Proof.
\end{IEEEproof}

\begin{figure}[tp]
\begin{center}
\psfrag{a}{{$(a)$}} \psfrag{b}{{$(b)$}} \psfrag{cs}{{$C'^s$}} \psfrag{cn}{{$C'^{t_4}$}} \psfrag{p}{\tiny{$\widehat{p}$}}
\psfrag{p1}{\tiny{$\widehat{p}_1$}} \psfrag{p2}{\tiny{$\widehat{p}_2$}}

\psfrag{s}{{{$s$}}} \psfrag{t1}{{{$t_1$}}} \psfrag{t4}{{{$t_4$}}}
\psfrag{t5}{{{$t_5$}}} \psfrag{t2}{{$t_2$}} \psfrag{t3}{{$t_3$}}
\psfrag{c1}{{\tiny{$\overline{C}^s_1$}}}
\psfrag{c2}{{\tiny{$\overline{C}^s_2$}}}
\psfrag{c2}{{\tiny{$\overline{C}^s_2$}}} \psfrag{t4}{{$t_4$}}
\psfrag{p*}{\tiny{$p^*$}} \psfrag{p13}{\tiny{$p_{\bar{1}3}$}}
\psfrag{12}{{$\perp_{12}$}} \psfrag{23}{{$\perp_{23}$}}
\psfrag{ab}{\tiny{$\ell^{ \bar{12}}_{\angle}$}}
\psfrag{13}{{$\perp_{13}$}}
\psfrag{15}{{$\perp_{15}$}}\psfrag{s13}{\tiny{$p^s_{\bar{1}3}$}}
\psfrag{lst}{{$\ell_{st_3}$}}
\includegraphics[width=1\columnwidth]{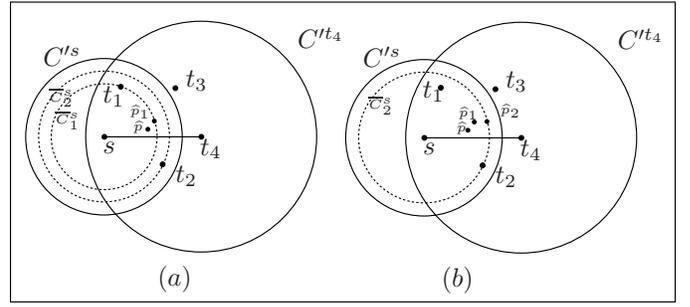}
\end{center}
\vspace{-4mm} \caption{{ $|T|=4$ example of
Figure~\ref{fig:Mincostalg1} is considered with
$\overline{T}=\{t_1,t_2\}$. (a): $\widehat{p}_1$ on the
circumference of dotted circle $\overline{C}_1$ is shown, which is
the optimal relay position solving $(s,T,\mathcal{Z},\gamma,F)$ in
the region $\overline{C}^s_2 \backslash \overline{C}^s_1$. (b) For
the region $C'^s \backslash \overline{C}^s_2$ and the optimal point
$\widehat{p}_2$ is shown on the circumference of $\overline{C}^s_2$.
}} \label{fig:Mincostalg2} \vspace{-6mm}
\end{figure}

\end{document}